\documentclass[%
 reprint,
superscriptaddress,
 amsmath,amssymb,
 aps,
floatfix,
]{revtex4-2}

\usepackage{xcolor}
\usepackage{svg}
\usepackage{multirow}
\usepackage{braket}
\usepackage{adjustbox}
\usepackage{makecell}
\usepackage{lipsum}
\usepackage{tabularray}
\usepackage{chngcntr}
\usepackage{longtable}
\usepackage{placeins}
\usepackage[normalem]{ulem}

\usepackage{graphicx}
\usepackage{dcolumn}
\usepackage{bm}

\usepackage{txfonts}

\newcommand{\comment}[1]{}
\begin{document}

\preprint{APS/123-QED}

\title{Macroscopic quantum entanglement between an optomechanical cavity and a continuous field in the presence of non-Markovian noise}

\author{S. Direkci}
    \email{sdirekci@caltech.edu}
    \affiliation{Theoretical Astrophysics 350-17, California Institute of Technology, Pasadena, California 91125, USA} 
\author{K. Winkler}
    \affiliation{Vienna Center for Quantum Science and Technology (VCQ), Faculty of Physics \& Vienna Doctoral School in Physics, University of Vienna, A-1090 Vienna, Austria}
\author{C. Gut}
    \affiliation{Vienna Center for Quantum Science and Technology (VCQ), Faculty of Physics \& Vienna Doctoral School in Physics, University of Vienna, A-1090 Vienna, Austria}
\author{K. Hammerer}
    \affiliation{Institute for Theoretical Physics and Institute for Gravitational Physics (Albert-Einstein-Institute),
Leibniz University Hannover, Appelstrasse 2, 30167 Hannover, Germany}    
\author{M. Aspelmeyer}
    \affiliation{Vienna Center for Quantum Science and Technology (VCQ), Faculty of Physics, University of Vienna, A-1090 Vienna, Austria}
    \affiliation{Institute for Quantum Optics and Quantum Information (IQOQI) Vienna, Austrian Academy of Sciences, Boltzmanngasse 3, 1090 Vienna, Austria} 
\author{Y. Chen}
    \affiliation{Theoretical Astrophysics 350-17, California Institute of Technology, Pasadena, California 91125, USA}


\date{\today}

\begin{abstract}

Probing quantum entanglement with macroscopic objects allows us to test quantum mechanics in new regimes. One way to realize such behavior is to couple a macroscopic mechanical oscillator to a continuous light field via radiation pressure. In view of this, the system that is discussed comprises an optomechanical cavity driven by a coherent optical field in the unresolved sideband regime where we assume Gaussian states and dynamics. We develop a framework to quantify the amount of entanglement in the system numerically. Different from previous work, we treat non-Markovian noise and take into account both the continuous optical field and the cavity mode. We apply our framework to the case of the Advanced Laser Interferometer Gravitational-Wave Observatory and discuss the parameter regimes where entanglement exists, even in the presence of quantum and classical noises.

\end{abstract}

\maketitle



\section{Introduction}

Entanglement is one of the hallmarks of the ``quantumness" of physical systems. Ideally, it is possible for macroscopic objects, massive and/or containing a high number of degrees of freedom, to be entangled with each other. Yet in practice, such macroscopic entanglement can be very delicate in the presence of decoherence.  It is an intriguing challenge to create and verify macroscopic entanglement, which is often viewed as expanding the limits of the quantum regime.  

Optomechanical systems are promising candidates for experimental demonstration of macroscopic entanglement, partly due to their theoretical robustness against mechanical decoherence imposed by coupling to a possibly highly populated thermal bath \cite{GENES200933}. They can also be used to engineer the quantum state of the mechanical system \cite{quantum_control}, where the entanglement is generated by the momentum exchange between the light reflecting from the mechanical oscillator---a phenomenon known as radiation pressure. It is theoretically well understood and broadly discussed in the literature
\cite{PhysRevA.78.032316, Miao2010, PhysRevA.81.012114, universal_quantum_entanglement, Danilishin2019,
Quantum_entanglement_teleportation_pulsed}; see for example, Refs. \cite{GENES200933, Aspelmeyer_cavity_optomechanics} for a review.

Entanglement in optomechanical devices has been widely studied and there have been several successful experimental realizations: stationary entanglement between simultaneous light tones mediated by an optomechanical device \cite{entangled_radiation_from_micromechanical,
propagating_optical_modes}, generation of entanglement between spaced mechanical oscillators both in the micro and macro regime via radiation pressure \cite{Remote_between_micromechanical,
stabilized_entanglement_massive, Direct_observation_of_macroscopic_entanglement, macroscopic_mechanical_and_spin}, and optomechanical entanglement between the light field and the mechanical oscillator in a pulsed scheme \cite{Mechanical_Motion_with_Microwave_Fields} are examples of such demonstrations. There also exist many proposals in the literature to further study macroscopic
quantum phenomena in optomechanical systems \cite{PhysRevA.84.052121, entanglement_macroscopic, EPR_entangled_motion_of, Macroscopic_quantum_entanglement_modulated,Pirandola_Entanglement_Swapping} and entanglement between coupled oscillators in the presence of non-Markovian baths \cite{ludwig_2010}. In this work, we consider stationary
optomechanical entanglement, where the system parameters (e.g. the driving), and statistical behavior thereof, are not changing over time. Schemes to verify stationary optomechanical entanglement were proposed in \cite{universal_quantum_entanglement, Miao2010, corentin_klemens_stat_optomech_entanglement, PhysRevLett.98.030405, PhysRevLett.99.250401}, whereas to the best of our
knowledge an experimental demonstration has not been performed yet. 

Our system consists of a single mechanical mode interacting with an optical cavity mode and the quadratures of the light field exiting the cavity. 
At any time $t$, we study the bipartite entanglement that is present in the joint quantum state between mechanical mode, optical mode, and the light that has exited the system during $t'\le t$. See Fig.~1 of Ref.~\cite{universal_quantum_entanglement}, which includes a space-time diagram that illustrates the configuration. 
In the regime where the dynamics are linear, the state is Gaussian,
and the noise processes are Markovian (white), the open-system optomechanical dynamics is solvable analytically and the state of the system can be known exactly \cite{PhysRevA.78.032316}. The white-noise model describes well devices with high-frequency oscillators, where only thermal excitations are expected, and in the limit of large bath temperature where $ k_B T \gg \hbar \omega_m$, $T$ is the temperature of the bath and $\omega_m$ is the resonance frequency of the oscillator \cite{PhysRevA.63.023812}. 

In this work, we extend the description to non-Markovian Gaussian noise  processes where analytical results are, to our knowledge, not available, thus requiring numerical methods. This approach is applicable whenever non-white-noise processes, such as structural damping \cite{structural_damping, structural_thermal_mirror,Evidence_for_structural_damping, non_markovian_micromech_brownian}, are relevant.
We extend the methods developed in \cite{universal_quantum_entanglement}, incorporating a cavity, and, more importantly, non-Markovian noise processes. The technique consists in computing the minimal symplectic eigenvalue of the partially transposed covariance matrix of the system, constructed with numerical methods, which provides a measure of appropriate bipartite entanglement.

We first investigate entanglement in a generalized setting, considering  a heavy suspended oscillator with a low mechanical resonance frequency. This corresponds to the free-mass limit, where the mechanical resonance frequency is much smaller than the other characteristic frequencies of the system. We work in this limit to study the general behavior of heavy suspended oscillators affected by environmental decoherence, however the algorithm does not require this assumption. Subsequently, we stop working in the free-mass limit and focus our attention on the Advanced
Laser Interferometer Gravitational-Wave Observatory (aLIGO), \cite{adv_ligo} using it as a case study. It has been recently shown that by injecting squeezed vacuum, the detector's quantum noise can in principle surpass the free-mass standard quantum limit (SQL) by 3\,dB ~\cite{correlations_kilogram_mass_mirrors}.

\begin{figure}
\includegraphics[width=0.7\columnwidth]{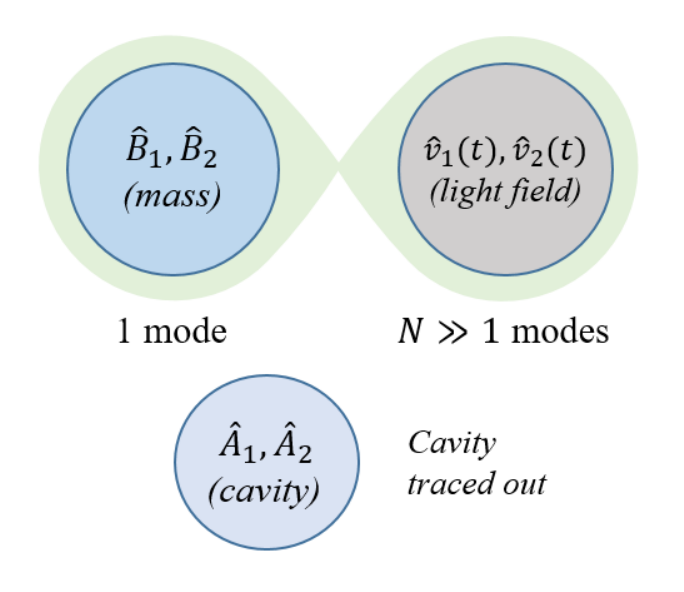}
\includegraphics[width=0.7\columnwidth]{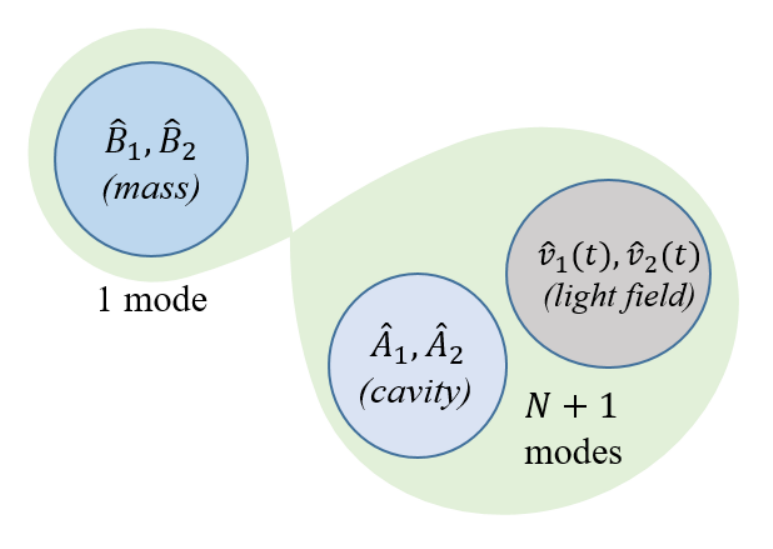}
\caption{\label{fig:partitions} Figurative representation of the two different partitions that is used while testing for entanglement, which are partitioning by tracing over (top row) and \textit{not} tracing over (bottom row) the cavity. Note that the system configuration is not changed, i.e. cavity is still present for both partitions.}
\end{figure}

It is natural to ask whether this can already imply that aLIGO has built quantum entanglement between the mirrors and the light field.  The answer to this question is non-trivial.  First, from ~\cite{correlations_kilogram_mass_mirrors}, we see that the level of classical noise is not yet below the SQL ~\footnote{The quantum noise's beating of the SQL is only inferred by subtracting a classical noise floor that was obtained through calibration}.  Second, the strict definition of entanglement we use here requires integrating over all frequencies: it remains uncertain whether having noise below the SQL within a certain finite frequency band automatically leads to entanglement. Therefore, we parametrize aLIGO's noise curves to investigate regimes where entanglement, according to its strict definition, exists. 

This paper is organized as follows: In Sec. \ref{sec:system_dynamics}, we introduce the dynamics of the system and its equations of motion. In Sec. \ref{sec:entanglement_criteria_partitions}, we state our entanglement criterion and the covariance matrix of the system for two partitions of interest. To show the usefulness of our technique for systems with low mechanical resonance frequencies, we investigate entanglement in a generalized setting 
in Sec. \ref{sec:non_markovian_general}. In Sec. \ref{sec:noise_model}, we give details about aLIGO's noise budget, and talk about how we model it in our calculations.
Finally, in Sec. \ref{sec:results}, we investigate whether there is entanglement between the mechanical oscillator and the light field at aLIGO for the partitions of interest, given different parametrizations of the classical noise curves.
\vspace{-1ex}
\section{System Dynamics}
\label{sec:system_dynamics}
\vspace{-2ex}
Let us consider an optical cavity with a movable mirror, driven by a laser with frequency $\omega_0$ close to one of the resonant frequencies of the cavity, $\omega_0+\Delta$ \cite{scaling_law}. 
The quantity $\Delta$ is often referred to as the detuning frequency of the cavity. For such a system, the linearized Hamiltonian in the interaction picture with the rotating-wave approximation (RWA) is given by \cite{Chen_2013}

\begin{equation}
\begin{gathered}
\label{eq:first_hamiltonian}
H = \hbar \omega_m \hat{B}^\dagger \hat{B} + \hbar \Delta \hat{A}^\dagger \hat{A} - \hbar G \hat{x} (\hat{A}^\dagger + \hat{A})      \\ + i \hbar \sqrt{2 \gamma}
    \int_{-\infty}^{\infty} \frac{d \Omega}{2 \pi} \left[\hat{A}^\dagger \hat{c}(\omega_0+\Omega) - \hat{A} \hat{c}^\dagger(\omega_0+\Omega) \right]\\ 
    + \int_{- \infty}^{\infty} \frac{d \Omega}{2 \pi} \left[\hbar \Omega \hat{c}^\dagger(\omega_0+\Omega) \hat{c}(\omega_0+\Omega)\right] \,,
\end{gathered}
\end{equation}
where $\hat{B}$ and $\hat{B}^\dagger$ are the annihilation and creation operators of the mechanical mode (center of mass motion of the mirror), $\hat{x}$ is the position of the center of mass of the mirror, $\omega_m$ is the mechanical resonance frequency, $\hat{A}$ and $\hat{A}^\dagger$ are the annihilation and creation operators of the cavity mode, $\hat{c}(\omega_0+\Omega)$ and $\hat{c}^\dagger(\omega_0+\Omega)$ are the annihilation and creation operators of the external vacuum light field at frequency $\omega_0+\Omega$, $G$ is the linear optomechanical coupling constant, and $\gamma$ is the decay rate of the cavity mode. The position and momentum operators of the mirror are related to the creation and annihilation operators of the mechanical mode by
\begin{subequations}
\begin{align}
\relax \hat{x} &= \sqrt{\frac{\hbar}{ M \omega_m}}  \, \frac{\left( \hat{B}+ \hat{B}^\dagger \right)}{\sqrt{2}} \,,\\
\relax \hat{p} &= \sqrt{\hbar M \omega_m} \, \frac{\left( \hat{B} - \hat{B}^\dagger \right)}{\sqrt{2}i} \,.
\end{align}
\end{subequations}
Note that for the sake of convenience we chose a displaced frame where all operators have zero mean. The mode operators satisfy the canonical commutation relations, 
\begin{equation}
\begin{gathered}
\relax [ \hat{A}, \hat{A}^\dagger ] = [ \hat{B}, \hat{B}^\dagger ]= 1.
\end{gathered}
\end{equation}

aLIGO detectors are power- and signal-recycled Fabry-Perot Michelson interferometers, which contain a high number of degrees of freedom.  However, the core optomechanics can still be studied by the Hamiltonian given above; this reduction manifests itself in the ``scaling-law'' relations governing aLIGO's sensitivity as parameters of the signal-recycling cavity are modified \cite{scaling_law}.  
From the scaling-law, the coupling constant $G$ is related to the parameters of the interferometer by,
\begin{equation}
\begin{gathered}
G = \sqrt{\frac{2 \omega_0 P_c}{ \hbar L c}} \,,
\end{gathered}
\end{equation}
where $L$ is the arm length of the interferometer (i.e. the cavity length), $P_c$ is the power circulating inside the cavity, and $c$ is the speed of light in vacuum.

We can transform the Hamiltonian such that the cavity mode $(A,A^\dagger)$ couples with the traveling wave at $z=0$ (where the point-wise cavity interface is located). We derive this transformation in Appendix \ref{app:travelling_wave}. We use $\hat u$ and $\hat v$ to label the field right before entering and right after exiting the cavity, respectively.

In this paper, we 
restrict ourselves to $\Delta=0$. In this resonant case, the system is unconditionally stable and it reaches a steady state, in which the Heisenberg equations can be solved using Fourier transformation ~\footnote{We use the convention $\mathcal{F}\{f(t)\} = \int_{-\infty}^{\infty} f(t)e^{i\omega t} dt$}.
To write down and solve the Heisenberg equations, instead of annihilation and creation operators we use the Caves-Schumaker quadrature operators \cite{caves_two_photon, caves_two_photon_2}:
\begin{subequations}
\label{Eq:caves}
\begin{align}
\hat{u}_1(\Omega) = \frac{\hat{u}(\omega_0+\Omega)+\hat{u}^\dagger(\omega_0-\Omega)}{\sqrt{2}} \,,\\
\hat{u}_2(\Omega) = \frac{\hat{u}(\omega_0+\Omega)-\hat{u}^\dagger(\omega_0-\Omega)}{\sqrt{2}i} \,,
\end{align}
\end{subequations}
where $\hat u_j^\dagger(\Omega) = \hat u_j(-\Omega)$. Quadratures $\hat{v}_1(\Omega)$ and $\hat{v}_2(\Omega)$ are defined from $\hat{v}(\omega_0+\Omega)$ and $\hat{v}(\omega_0-\Omega)$ in a similar fashion. Their commutation relations are \cite{Danilishin_quantum_meas_theory}
\begin{subequations}
\begin{align}
[\hat{u}_1(\Omega),\hat{u}_2(\Omega')] &= [\hat{v}_1(\Omega),\hat{v}_2(\Omega')] =  2 \pi \delta(\Omega+\Omega') \,,\\
[\hat{u}_j(\Omega),\hat{u}_j(\Omega')] &= [\hat{v}_j(\Omega), \hat{v}_j(\Omega')] = 0 \,,
\end{align}
\end{subequations}
for $j =1,2$. Then, in the time domain, we have
\begin{subequations}
\begin{align}
\hat{u}_j(t) &= \int_{-\infty}^{\infty} \frac{d \Omega}{2\pi} \hat{u}_j(\Omega) e^{-i \Omega t} \,, \\
[\hat{u}_{1}(t),\hat{u}_{2}(t')] &= i \delta(t-t') \,, \\
[\hat{u}_{1}(t),\hat{u}_{1}(t')] &= [\hat{u}_{2}(t),\hat{u}_{2}(t')] = 0 \,,
\end{align}
\end{subequations}
for $j=1,2$.  We similarly define quadrature operators $\hat A_{1,2}$ and $\hat B_{1,2}$, in the {\it time domain}, with 
\begin{equation}
    \hat A_1(t)=\frac{\hat A(t) + \hat A^\dagger(t)}{\sqrt{2}}
\,,\quad 
    \hat A_2(t)=\frac{\hat A(t) - \hat A^\dagger(t)}{\sqrt{2}i}\,,
    \label{Aquadratures}
\end{equation}
and similarly for $\hat B_{1,2}$.   We also have
\begin{align}
[\hat A_1(t), \hat A_1(t)]&=[\hat A_2(t), \hat A_2(t)]=0\,,\\ 
[ \hat A_1(t), \hat A_2(t)]&=i\,,
\end{align}
and the same for $\hat B_{1,2}$.
Note here that the commutators are for same-time operators. 

We include two classes of ``classical" noises \cite{Danilishin_quantum_meas_theory} in our system: a force noise $\hat{n}_F$ and a sensing noise $\hat{n}_X$, originally arising from a quantum treatment of the interaction of the system with its environment. We label them as ``classical" noises because we assume that the details -- and possible quantum limit -- of a microscopic model of these noises are irrelevant (typically because they arise from thermal baths in the high temperature limit) unlike the noise arising from vacuum fluctuations, $\hat{u}_1(\Omega)$ and $\hat{u}_2(\Omega)$. Force noise affects the center-of-mass motion of the mechanical oscillator by introducing fluctuations in its momentum. 
We also introduce a velocity damping of the oscillator, with a damping rate $\gamma_m$. $\gamma_m$ and $n_F$ are associated with the heat bath(s) the mass is coupled to, with the value of $\gamma_m$ and the spectrum of $n_F$ related by the fluctuation-dissipation theorem ~\cite{kubo1966fluctuation}.  Sensing noise affects how the position is measured by the light field. In our model below it arises from fluctuations of the reflecting surface that introduces noise in the cavity field. 

In the Heisenberg picture, the dynamics are given by the Langevin equations of motion. In the Fourier domain they are written as
\begin{subequations}
\label{Eq:final_Eqs}
\begin{align}
-i\Omega \hat A_1& =-\gamma \hat A_1 +\sqrt{2\gamma}\hat u_1 \,,
\label{eqA1}
\\
-i\Omega \hat A_2 &=-\gamma \hat A_2+ \sqrt{2\gamma}\hat u_2 + \sqrt{2} G(\hat x +\hat n_X) \,,
\label{eqA2}
\\
\label{eqX}
-i\Omega \hat x &= \hat p/M \,, \\
-i\Omega \hat p & = -\gamma_m \hat p -M\omega_m^2\hat x +\sqrt{2}\hbar G \hat A_1 + \hat n_F \,,
\label{eqP}
\\
\label{eqv1}
\hat v_1 &= \hat u_1 -\sqrt{2\gamma} \hat A_1 \,,\\
\label{eqv2}
\hat v_2 &= \hat u_2 -\sqrt{2\gamma} \hat A_2 \,. 
\end{align}
\end{subequations}
We refer to Eqs.~\eqref{Eq:final_Eqs} as the Heisenberg equations for the rest of the article. It is straightforward to solve them to obtain $(\hat x, \hat p, \hat A_{1,2}, \hat v_{1,2})$ in terms of the input fields, $(\hat u_{1,2},\hat n_X,\hat n_F)$, 
referred as the input-output relations of the system.  
More specifically, quantum fluctuations in the in-going quadratures $\hat{u}_{1,2}(\Omega)$ drive the system's quantum noise~\cite{Kimble_conversion_of_conventional}. From Eq.~\eqref{eqv1} and \eqref{eqv2}, reading the out-going field quadratures are subject to noises in $\hat u_1$ and $\hat u_2$, giving rise to the {\it shot noise} (SN) for that readout strategy~\cite{Kimble_conversion_of_conventional}. On the other hand, from Eq.~\eqref{eqA1}, we see that $\hat u_1$ drives $\hat A_1$, which in Eq.~\eqref{eqP} drives the momentum of the test mass, which then shows up in the position of the test mass via Eq.~\eqref{eqX}, giving rise to {\it quantum radiation pressure noise} (QRPN), also known as backaction noise in the literature. In general, the power spectrum of the SN is inversely proportional to circulating power in the cavity, while that of the QRPN is proportional to circulating power. 
\vspace{-1ex}
\section{Entanglement Criteria and Partitions}
\label{sec:entanglement_criteria_partitions}
\vspace{-2ex}
The canonical commutation relations imply that $\mathbf{V} + \frac{1}{2} \mathbf{K}$ is positive semidefinite, where $\mathbf{V}$ is the covariance matrix with $\mathbf{V}_{ij} = \langle \{\hat{X}_i - \langle \hat{X}_i \rangle, \hat{X}_j - \langle \hat{X}_j \rangle \} \rangle/2 $ and $\mathbf{K}_{ij} = [ \hat{X}_i, \hat{X}_j ] $ is the commutator matrix of the quadratures in the system. This relation can be stated as
\begin{equation}\label{Eq:Heisenberg_uncertainty}
    \mathbf{V} + \frac{1}{2} \mathbf{K} \geq 0 \,.
\end{equation}
Here for an $N$-partite system containing $N$ harmonic oscillators, the matrices $\mathbf{V}$ and $\mathbf{K}$ are $2N$-dimensional.

To test for bipartite entanglement in a multimode system, we use 
the positivity of the partial transpose (PPT) criterion, which is necessary and sufficient to test for the separability of one of the modes from the rest for Gaussian systems \cite{peres_ppt, Adesso_2007, bound_entangled_states}.

To use the PPT criterion in this context, one obtains the partial-transposed covariance matrix $\mathbf{V}_{\rm pt}$ by reverting the momentum of that one mode (which puts a minus sign on the column and the row that contains the momentum in question) \cite{simon}. The PPT criterion for separability is expressed as
\begin{equation}
\label{Eq:ppt}
    \mathbf{V}_{\rm pt} + \frac{1}{2} \mathbf{K} \geq 0 \Leftrightarrow {\rm Separability} \,.
\end{equation}

The amount of entanglement is quantified by the logarithmic negativity, $E_\mathcal{N}$ \cite{negativity}. For a Gaussian state of N modes, it is defined as
\begin{equation}
\label{Eq:negativity}
E_\mathcal{N} = \sum_{j=1}^{N} \text{max}\{0, -\text{log}_2 (\tilde{\nu}_j)\} \,,
\end{equation}
where $\tilde{\nu}_j$, $j=1, \dots N$ are the symplectic eigenvalues of the partially transposed covariance matrix, $\mathbf{V}_{\rm pt}$, which are given by the absolute values of the eigenvalues of $\mathbf{K}^{-1}\mathbf{V}_{\rm pt}$. For 1 vs. $N-1$ mode partitions, only one of the symplectic eigenvalues of $\mathbf{V}_{\rm pt}$ can have a magnitude smaller than 1 \cite{Serafini2017}, therefore there can be at most one negative eigenvalue of $\mathbf{V}_{\rm pt} + \frac{1}{2} \mathbf{K}$. We label the corresponding symplectic eigenvalue as $\tilde{\nu}_{min}$.




Using the PPT criterion, we test for entanglement between the mechanical oscillator and the optical field in two ways: first, we construct the covariance matrix $\mathbf{V}$ with the mechanical mode and the modes of the light field, essentially tracing out the cavity mode. Here, we perform the partial transpose operation with respect to the mechanical oscillator. Second, we include the cavity mode in the covariance matrix while still taking the partial transpose with respect to the mechanical oscillator, which corresponds to measuring the entanglement between the oscillator and the joint system of the cavity plus external light field. The two ways of partitioning are depicted in Fig. \ref{fig:partitions}. The elements of the covariance matrix $\mathbf{V}$ for both types of partitions, as well as the discretization of $\mathbf{V}$, can be found in Appendix \ref{app:cov_matrix}.

\vspace{-1ex}
\section{Entanglement in the presence of Non-Markovian Noises}
\label{sec:non_markovian_general}
\vspace{-2ex}
Due to the numerical nature of the algorithm, we can tackle any noise spectral density associated with $\hat{u}_1(\Omega)$, $\hat{u}_2(\Omega)$, $\hat{n}_F(\Omega)$ and $\hat{n}_X(\Omega)$ using the PPT criterion defined in Eq. (\ref{Eq:ppt}) to determine whether entanglement is present for a given partition. Conversely, this problem is analytically solvable only for some simplified noise models to our knowledge, such as assuming all the noise sources to have a white spectrum \cite{universal_quantum_entanglement, corentin_klemens_stat_optomech_entanglement}.

To show the usefulness of the method, we investigate entanglement in heavy suspended oscillators with relatively low mechanical resonance frequencies. Examples of such systems are aLIGO, KAGRA \cite{KAGRA}, and VIRGO \cite{VIRGO}, but also smaller devices such as those in \cite{thomas_corbitt, komori_attonewton_2020}. For such systems, the mechanical resonance frequency, $\omega_m$, is much smaller than the other frequencies of the system, which is referred to as the \textit{free-mass limit}. In this setting, $\omega_m$ essentially does not affect the dynamics. Furthermore, in this limit where $ \Omega \gg \omega_m, \gamma_m$, the tradeoff between shot noise and QRPN gives rise to the SQL \cite{braginsky_khalili_thorne_1992}, given by
\begin{equation}
\label{Eq:sql_free_mass}
S_{\rm SQL}(\Omega) = \frac{2 \hbar}{M\Omega^2}.
\end{equation}

In the context of suspended oscillators, $\hat{n}_F(\Omega)$ is the force that gives rise to the suspension thermal noise, whereas $\hat{n}_X(\Omega)$ is an effective displacement that gives rise to coating thermal noise. When thermal noise is due to the internal friction of the suspension or the oscillator, the noise spectrum of $\hat{n}_F(\Omega)$ and $\hat{n}_X(\Omega)$ decreases as $1/\Omega$ above internal resonances, which is referred to as \textit{structural damping} \cite{structural_damping, torsion_pendulum, structural_thermal_mirror}. Evidently, structural damping gives rise to non-Markovian noises, and the position-referred noise spectral densities of $\hat{n}_F(\Omega)$ and $\hat{n}_X(\Omega)$ are given, in the free-mass limit, by
\begin{subequations}
\label{Eq:general_spectra}
\begin{align}
S_{F}(\Omega) &= \frac{2\hbar}{M} \frac{\Omega_F^3}{|\Omega|^5} \,, \\
S_{X}(\Omega) &= \frac{2\hbar}{M} \frac{1}{\Omega_X |\Omega|} \,,
\end{align}
\end{subequations}
where $\Omega_F$ and $\Omega_X$ are the frequencies where the respective noise curves cross the SQL, given in Eq. (\ref{Eq:sql_free_mass}). Accordingly, they encode the strength of the noise processes $n_F$ and $n_X$, relative to the SQL level. For $\hat{n}_F(\Omega)$, the position-referred spectrum is related to the noise spectral density, labeled as $S_{n_F}(\Omega)$, with $S_{F}(\Omega) = S_{n_F}(\Omega)/M^2\Omega^4$, whereas for $\hat{n}_X(\Omega)$, the position-referred spectrum $S_{X}(\Omega)$ is also the noise spectral density $S_{n_X}(\Omega)$ \footnote{To find spectra of the form of Eqs.~\eqref{Eq:general_spectra}, one assumes the free-mass and high-$Q$ limits where the mechanical susceptibility behaves as $1/\Omega^2$; additionally, one must be in frequency ranges where $S_{n_F}$ and $S_{n_X}$ are constant polynomial roll-off -- typical at frequencies above the resonance of suspended oscillators.}. The incoming field quadratures $\hat u_j$ have uncorrelated white spectra given by Eq.~\eqref{Eq:vacuum_spectra}, since we assume the incoming field to be at vacuum state.

In the limit of large cavity bandwidth, $\gamma \gg \Omega$ the cavity can be eliminated adiabatically. Then, the equations of motion in (\ref{Eq:final_Eqs}) are modified as
\begin{subequations}
 \label{Eq:Eqs_without_cavity}
\begin{align}
\hat{v}_1(\Omega) &= \hat{u}_1(\Omega) \,, \\
\hat{v}_2(\Omega) &= \hat{u}_2(\Omega) + \alpha (\hat{x}(\Omega)+\hat{n}_X(\Omega)) \,, \\
\begin{split}
-i\Omega \hat{p}(\Omega) &= -\gamma_m \hat{p}(\Omega) -M\omega_m^2 \hat{x}(\Omega) \\ &+\hbar \alpha \hat{u}_1(\Omega) +\hat{n}_F(\Omega) \,,
\end{split}\\
-i\Omega \hat{x}(\Omega) &= \hat{p}(\Omega)/M \,,
\end{align}
\end{subequations}
where $\alpha = \Omega_q \sqrt{M/\hbar}$ and $\Omega_q = 2G \sqrt{\hbar/M\gamma}$ is the characteristic interaction frequency. In the context of structural damping, there is no velocity damping, Instead, the damping arises from a complex spring constant associated with the mechanical oscillator. Accordingly, Eq. (\ref{Eq:Eqs_without_cavity}b) is modified as
\begin{equation}
\begin{split}
-i\Omega \hat{p}(\Omega) &= -M\omega_m^2(1+i \, \phi(\Omega)) \hat{x}(\Omega) \\ &+\hbar \alpha \hat{u}_1(\Omega) +\hat{n}_F(\Omega) \,,
\end{split}
\end{equation}
where $\phi(\Omega)$ is referred to as the \textit{loss angle}. When structural damping is present, $\phi(\Omega)$ is constant for a large band of frequencies and goes to zero as $\Omega \rightarrow 0$, however the dependence of $\phi(\Omega)$ to $\Omega$ depends on the properties of the material \cite{structural_damping}. Numerically, we choose to model this with $\phi(\Omega) = \phi \cdot \Omega/(\Omega + \Omega_c) $, so that $\phi(\Omega) \approx \phi$ for $ \Omega \gg \Omega_c$ and $\phi(0) = 0 $ for some cutoff frequency $\Omega_c$. Then, $\Omega_c$ determines the noise power of $\hat n_F$ and $\hat n_X$ at 0 Hz.

\begin{figure}
    \centering
    \includegraphics[width=\columnwidth]{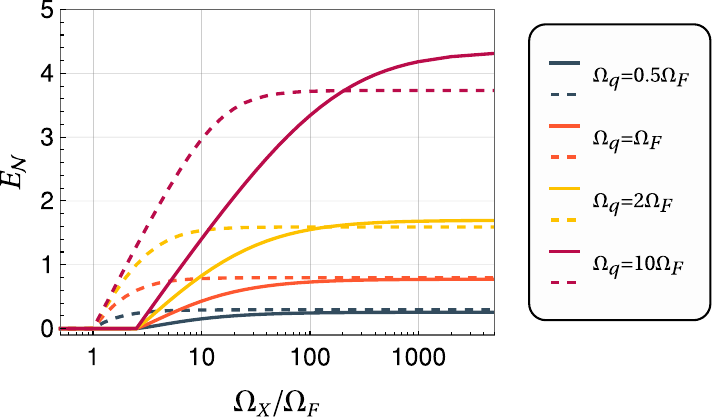}
    \caption{Logarithmic negativity between the mechanical oscillator and the outgoing light field in the free mass limit as a function of $\Omega_X/\Omega_F$ for various $\Omega_q/\Omega_F$. We plot the results for Markovian and non-Markovian force and sensing noise with dashed and plain lines, respectively. Note that entanglement does not exist for $\Omega_X/\Omega_F \lesssim 1 $ with Markovian and for $\Omega_X/\Omega_F \lesssim 2.6 $ with non-Markoivan noise sources, for $\omega_m / (2\pi) = 1 $ Hz, $\gamma_m / (2\pi) = 0.01 $ Hz, $\phi = 0.05 $, and a cutoff frequency of $\Omega_c/(2 \pi) = 0.05 $ Hz.}
    \label{fig:omega_q_to_omega_f}
\end{figure}

We fix $\omega_m / (2\pi) = 1 $ Hz, $\Omega_F / (2\pi) = 100 $ Hz, $\Omega_c / (2\pi) = 0.05 $ Hz, $\phi = 0.05$, and vary $\Omega_X$ and $\Omega_q$. Note that the aim of these choices is to ensure that $\Omega_F \gg \omega_m$, i.e. that the free-mass limit is justified. This choice also implies the large bath temperature limit where $ k_B T \gg \hbar \omega_m$, $T$ being the temperature of the bath. Furthermore, $\Omega_q \gg \omega_m$ ensures that the measurement of the system performed by light is faster than the dynamics of the system. Lastly, we perform the simulations by sampling the covariance matrix for $\tau = 0.1$ s, from $t=-0.1$ s to t=0 s, where $t$ indicates the time (for the sampling of the covariance matrix, see Appendix \ref{app:cov_matrix}). Hence, we cannot resolve the very low frequency regime of $\Omega_c$. Physically, this corresponds to the finite detection frequency resolution that renders the behavior of $\phi$ at the cutoff inaccessible. In this sense, one might expect $\Omega_c$ to be irrelevant. However, we see that it does affect the final characterisation of entanglement, due to its contribution to the total variance of the quadratures of the mechanical oscillator.

After specifying the low-frequency behavior of $\phi$ with $\Omega_c$, the relevant parameters on which the presence of entanglement depends are $\Omega_q$, $\Omega_F$, and $\Omega_X$. Then, working in the free-mass limit enables us to examine the general behavior of suspended oscillators with low
mechanical resonance frequencies, classified by their coating materials and suspension systems (i.e. the low-frequency behavior of $\phi$). Therefore, we look for entanglement between the oscillator and the outgoing light field by varying the ratio $\Omega_X / \Omega_F$ for various $\Omega_q$ in Fig. \ref{fig:omega_q_to_omega_f}, and we plot the results with plain lines. We find that for $\Omega_c/(2\pi) = 0.05$ Hz, entanglement does not exist for $\Omega_X/\Omega_F \lesssim 2.6$ for any value of $\Omega_q$. For $\Omega_X/\Omega_F \gtrsim 2.6$, the system is entangled for any (finite) $\Omega_q$, and the entanglement increases monotonously with increasing $\Omega_X /\Omega_F$. This implies the existence of ``universal" entanglement, meaning that whether the system is entangled or not is independent of $\Omega_q$, the interaction frequency (or, in other words, how fast the system is measured by light). When the system is entangled, the amount of entanglement increases when $\Omega_q$ is increased. We also note that the threshold for $\Omega_X/\Omega_F$ above which the system is entangled depends on the low-frequency cutoff $\Omega_c$ chosen in our model for the spectra of $\hat n_F$ and $\hat n_X$. We saw that the threshold is inversely proportional to the cutoff frequency.

\begin{figure}
\includegraphics[width=0.9\columnwidth]{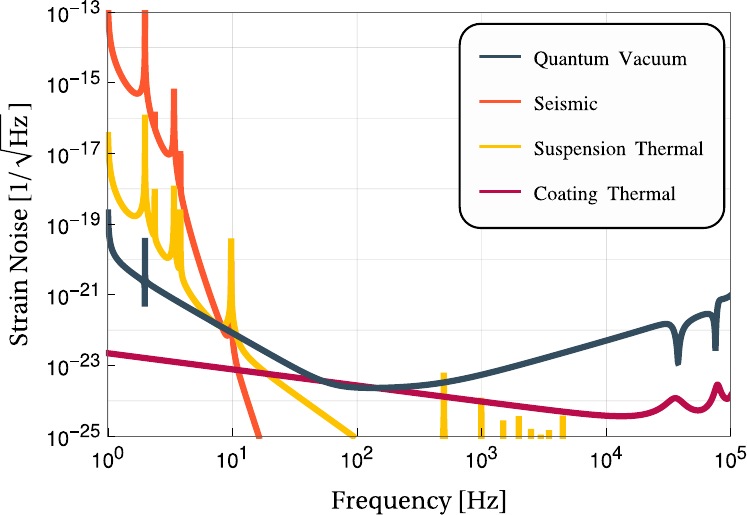}
\caption{\label{fig:aLIGO_noise_curves} aLIGO noise budget obtained from pygwinc. Only the dominant classical noise sources are plotted, along with the quantum noise. The total force noise in the system is the sum of the seismic noise and the suspension thermal noise, which is effective at low frequencies. The coating Brownian thermal noise is taken as the main constituent of the sensing noise. As can be seen from the figure, quantum noise dominates over the sensing noise by a large margin at high frequencies. }
\end{figure}

In order to see the significance of non-Markovianity on the results, we repeat the same procedure with a white force and sensing noise. The noise spectra are given by $S_{n_F}(\Omega) = 2\hbar M \Omega_F^2$, $S_{F}(\Omega) = 2\hbar \Omega_F^2 / M \Omega^4$, and $S_{n_X}(\Omega) = S_{X}(\Omega) = 2\hbar/M \Omega_X^2$. The results can again be found in Fig. \ref{fig:omega_q_to_omega_f}, plotted with dashed lines, where the system is not entangled for any $\Omega_q$ when $\Omega_X/\Omega_F \lesssim 1$, whereas for $\Omega_X/\Omega_F \gtrsim 1$, entanglement exists for all $\Omega_q$ and the amount of entanglement increases with increasing $\Omega_q$. Since we see this behavior for both Markovian and non-Markovian noise, we prove that the universality of the entangling-disentangling phase transition is independent of the power spectral densities of the classical noises, and that the power spectral densities only determine the threshold above which we have entanglement for all $\Omega_q$ in a manuscript that is currently in preparation.
\vspace{-1ex}
\section{Noise Model of \MakeLowercase{a}LIGO}
\label{sec:noise_model}
\vspace{-2ex}
The primary noise sources in aLIGO, other than the quantum noise, are the following \cite{adv_ligo}: seismic noise and suspension thermal noise are the main constituents of the force noise, and mirror coating thermal noise constitutes the sensing noise. The noise spectrum is dominated by seismic and thermal noise at low frequencies (until 100 Hz), and quantum noise at high frequencies, cf. Fig.~\ref{fig:aLIGO_noise_curves}. The interferometer noise is stationary and Gaussian to very good approximation in the absence of glitches (i.e. transient noise artifacts) \cite{Guide_to_LIGO_Virgo_detector_noise}.

Seismic noise occurs because of the ground motion at the interferometer sites. This motion is $ \sim 10^{-9} \text{m} / \sqrt{\text{Hz}}$ at 10 Hz \cite{Sensitivity_of_Advanced_LIGO}. To provide isolation from this motion, the mirrors are suspended from quadruple pendulums \cite{quadruple_pendulum}. The primary components of thermal noise are suspension thermal noise and coating Brownian noise. Suspension thermal noise occurs due to loss in the fused silica fibers used in the final suspension stage \cite{adv_ligo}, whereas the coating Brownian noise (which is classified as a sensing noise) occurs due to the mechanical dissipation in the coatings \cite{coating_thermal}. Other types of sensing noise comprise many noise sources that are dominant at high frequencies, such as thermal fluctuations of the mirror’s shape, optical losses, or photodetection inefficiency \cite{Muller_Ebhardt_quantum_state_preparation}. The noise budget of aLIGO can be found in Figure \ref{fig:aLIGO_noise_curves}.

\begin{figure}
\includegraphics[width=0.85\columnwidth]{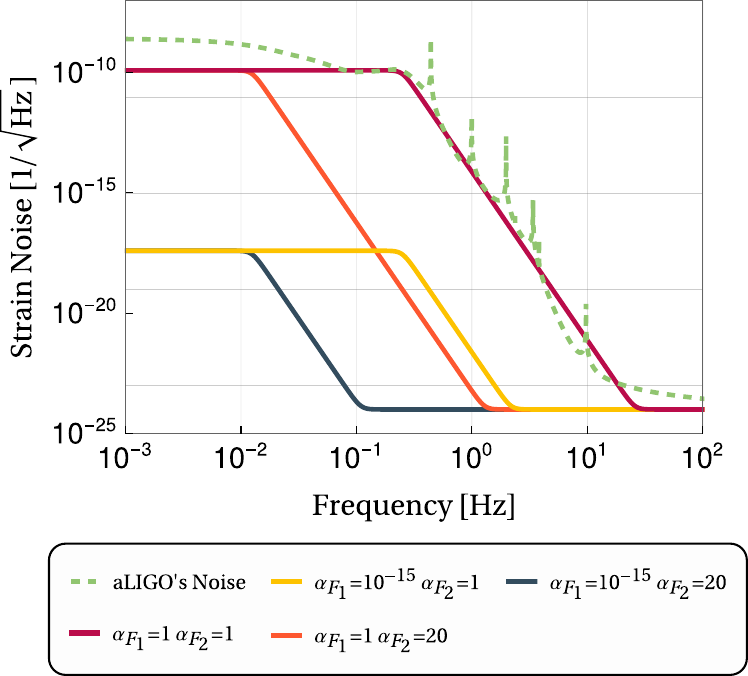} \\ \vspace{0.25cm}
\includegraphics[width=0.85\columnwidth]{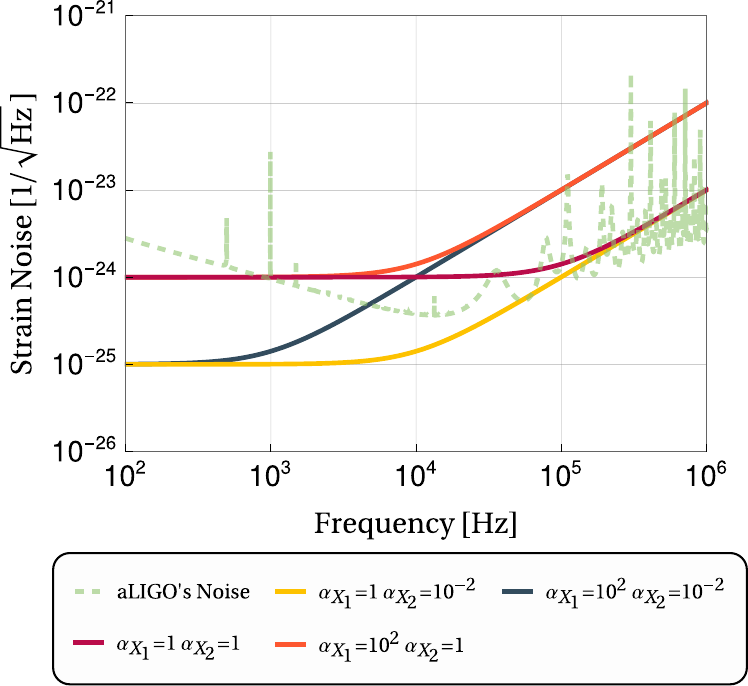}
\caption{\label{fig:force_and_sensing_noise_spectrum} Force (top row) and sensing noise (bottom row) spectra parametrized by $\alpha_{F_1}$, $\alpha_{F_2}$, $\alpha_{X_1}$, and $\alpha_{X_2}$. The effect of $\alpha_{F_1}$ is to rise and lower the nominal noise strength below the cutoff frequency and $\alpha_{F_2}$ shifts the cutoff frequency. Similarly, $\alpha_{X_1}$ shifts the cutoff frequency where the sensing noise starts increasing as $\Omega^2$, whereas $\alpha_{X_2}$ shifts the nominal noise level.}
\end{figure}

Due to the classification above, we represent the sum of the seismic and the suspension thermal noise with $\hat{n}_F(\Omega)$, and the coating thermal noise with $\hat{n}_X(\Omega)$. We use the aLIGO noise budgets as given by the Python Gravitational Wave Interferometer Noise Calculator library (pygwinc) \cite{pygwinc} and we model them with rational functions of $\Omega^2$. The noise spectra are modelled by

\begin{subequations}
\label{Eq:S^{LIGO}_{F}_s_x}
\begin{align}
S^{LIGO}_{F}(\Omega) &= \frac{\tau_F \alpha_{F_1}}{\left(\frac{\Omega}{\omega_F} \alpha_{F_2}\right)^{14}+1} \,, \\
S^{LIGO}_{X}(\Omega) &= \tau_{X_1} \left(\frac{\Omega}{\omega_X}\right)^2 \alpha_{X_1} + \tau_{X_2} \alpha_{X_2} \,,
\end{align}
\end{subequations}
where $S^{LIGO}_{F}(\Omega)$ is the spectrum of the force noise, and $S^{LIGO}_{X}(\Omega)$ is the spectrum of the sensing noise. We model $S^{LIGO}_{F}(\Omega)$ to decay as $\Omega^{-14}$ instead of $\Omega^{-16}$ (which is the expected behavior for quadruple suspension systems) since it performs better at approximating the global behavior. The power spectral densities are characterized by the time constants $\tau_F$, $\tau_{X_1}$, $\tau_{X_2}$ and cutoff frequencies $\omega_F$, $\omega_X$. The values of these parameters can be found in Appendix \ref{appendix:parameter_tables}, Table \ref{tab:noiseParams}. $\alpha_{F_1}$, $\alpha_{F_2}$, $\alpha_{X_1}$, and $\alpha_{X_2}$ are dimensionless constants that will be used to change the noise curves in Section \ref{sec:results}. Their effect on the noise curves can be seen in Fig. \ref{fig:force_and_sensing_noise_spectrum}. If we set all of them to be unity, we get our model of aLIGO noise curves.

Even though the parameters of aLIGO are well-known, they need to be recomputed since we are reducing the antisymmetric mode of the interferometer to a single cavity. The relation between the parameters of the antisymmetric mode and the parameters of the reduced cavity have already been computed \cite{scaling_law}, however, we choose to work numerically and find the parameters of the reduced cavity $G$, $\gamma$, $M$, $\gamma_m$, $\omega_m$, and $L$ by fitting aLIGO's quantum noise spectrum found in pygwinc. The modeled classical and quantum noise curves can be found in Appendix \ref{appendix:noise_model}, Fig. \ref{fig:total_noise_spectrum}. The fitted parameters can be found in Appendix \ref{appendix:parameter_tables}, Table \ref{tab:ligoParams}.
\vspace{-1ex}
\section{Entanglement in \MakeLowercase{a}LIGO}
\label{sec:results}
\vspace{-2ex}
In aLIGO, the spectrum is dominated by force noise for low frequencies and quantum noise for high frequencies, therefore we expect the sensing noise not to affect the entanglement significantly, which we observed with our numerics. Then, we focus on the effect of the force noise on the entanglement and use the parameters $\alpha_{F_1}$ and $\alpha_{F_2}$ to modify the force noise spectrum, and set $\alpha_{X_1} = \alpha_{X_2} = 1$ throughout all of the following subsections. We also introduce resonant modes to investigate the system as accurately as possible. Intuitively, we expect the entanglement to be destroyed in the presence of high classical noise levels.
\vspace{-2ex}
\subsection{Effect of Force Noise}
\label{sec:effect_of_force_noise}
\begin{figure}
\includegraphics[width=0.75\columnwidth]{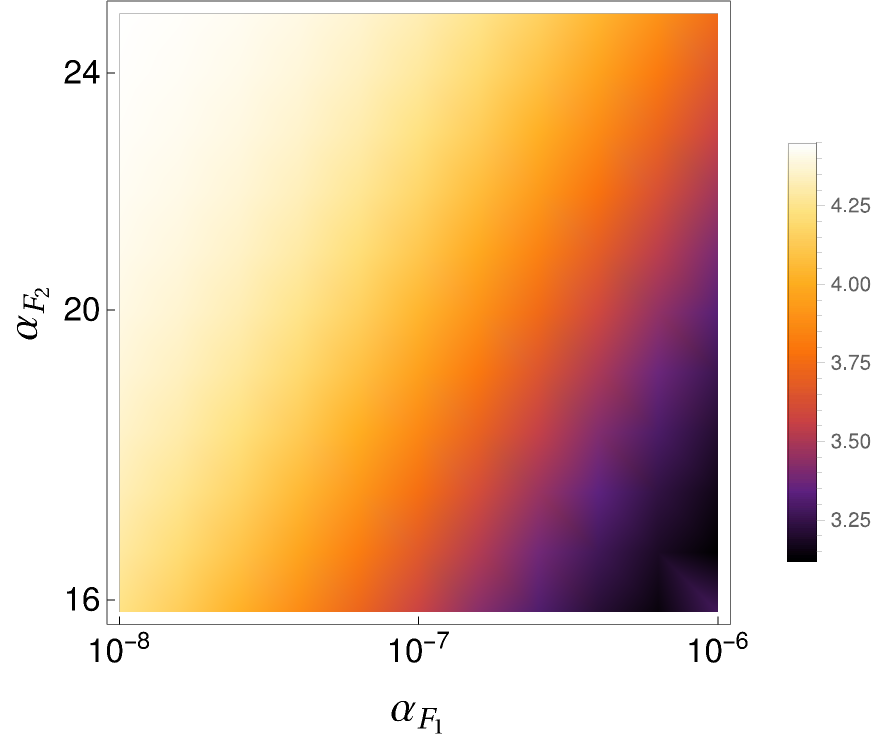}
\caption{\label{fig:contour_plots_symplectic} The effect of the force noise spectrum on the logarithmic negativity when we do not trace over the cavity. Note that the force noise levels increase toward the bottom-right of the figure and our fit of aLIGO's operation regime is for $\alpha_{F_1} = \alpha_{F_2} =1$.}
\end{figure}

\vspace{-2ex}
First, we investigate the effect of the force noise spectrum on entanglement and we calculate the logarithmic negativity $E_\mathcal{N}$ as a function of $\alpha_{F_1}$ and $\alpha_{F_2}$, for both of the partitions described in Sec. \ref{sec:entanglement_criteria_partitions}. For all pairs $\alpha_{F_1}$ and $\alpha_{F_2}$ here, we find larger logarithmic negativity values when we do not trace over the cavity. The results for the partition where we do not trace over the cavity can be found in Fig. \ref{fig:contour_plots_symplectic}. The amount of entanglement in the system diminishes when the force noise increases: that is, towards the bottom-right of the plot where $\alpha_{F_1}$ increases (proportional to the DC noise power) and $\alpha_{F_2}$ decreases (inversely proportional to the noise bandwidth), cf. Sec. \ref{sec:noise_model} and Fig. \ref{fig:force_and_sensing_noise_spectrum}. Our fit of aLIGO's force noise level is for $\alpha_{F_1} = \alpha_{F_2} =1$, hence this plot is for a comparatively low level of force noise. Further to the bottom-right, the numerics become unstable and do not converge due to the wide range of orders of magnitude entering the calculation; see Appendix \ref{sec:numerical_implementation} for a discussion of the numerical implementation. Therefore, we cannot give a definite answer about optomechanical entanglement in aLIGO with our model yet.

Next, we look for entanglement in the absence of the seismic noise, since it is the dominating contribution to the force noise of the system. Not being subjected to seismic noise is a realistic scenario for space-based gravitational-wave interferometers, such as the Laser Interferometer Space Antenna (LISA) \cite{spacebased}. For aLIGO, in the absence of seismic noise, suspension thermal noise dominates the spectrum for low frequencies, which is modelled as
\begin{equation}
\label{Eq:suspension_thermal}
S_{ST}(\Omega) = \frac{\tau_{ST}}{\left(\frac{\Omega}{\omega_{ST}} \right)^{8}+1} \,,
\end{equation}
similar to how the total force noise was modelled in Sec. \ref{sec:noise_model}. The parameters $\tau_{ST}$ and $\omega_{ST}$ can be found in Table \ref{tab:noiseParams}. Without changing the other parameters in the system, we find that we can achieve negativities of 1.52 and 1.72 for the partitions where we do and do not trace over the cavity, respectively. This means that, in the absence of seismic noise, aLIGO has stationary optomechanical entanglement in its current operating regime.
\vspace{-2ex}
\subsection{Effect of Low Frequency Resonances}
\vspace{-2ex}
Both the classical and the quantum noise curves contain many resonances, as can be seen from Fig. \ref{fig:aLIGO_noise_curves}. The resonances in the seismic and suspension thermal noise arise from the displacement noises of the rigid-body resonant modes of the 4-stage suspension system \cite{quadruple_pendulum}. These modes can be modeled with a sum of Lorentzians multiplying the force noise spectrum defined in Eq. (\ref{Eq:S^{LIGO}_{F}_s_x}). Then, the new formula defining the spectrum of the force noise is given as

\begin{subequations}
\label{Eq:S^{LIGO}_{F}_lorentzians}
\begin{align}
\begin{split}
S^R_{F}(\Omega) &= \frac{\tau_F \alpha_{F_1}}{\left(\frac{\Omega}{\omega_F} \alpha_{F_2}\right)^{14}+1} \Biggl( 1 + \\ 
& \sum_i \frac{A_{v_i}^2}{(\Omega - \Omega_{v_i})^2 + (\frac{1}{2} \Gamma_{v_i})^2} \Biggr) \quad \text{with}\ \Omega > 0 \,,
\end{split} \\
\begin{split}
S^R_{F}(\Omega) &= S^R_{F}(-\Omega) \,, \\
\end{split}
\end{align}
\end{subequations}
where the sum is over the resonant modes. The parameters $\Omega_{v_i}$, $\Gamma_{v_i}$, and $A_{v_i}$ are the mode frequencies, full widths at half maximum (FWHM), and the amplitudes of the Lorentzians, respectively, and they are listed in Appendix \ref{appendix:parameter_tables}, Table \ref{tab:violinModeParams} for the modes with the biggest relative amplitudes.

\begin{figure}
\includegraphics[width=0.8\columnwidth]{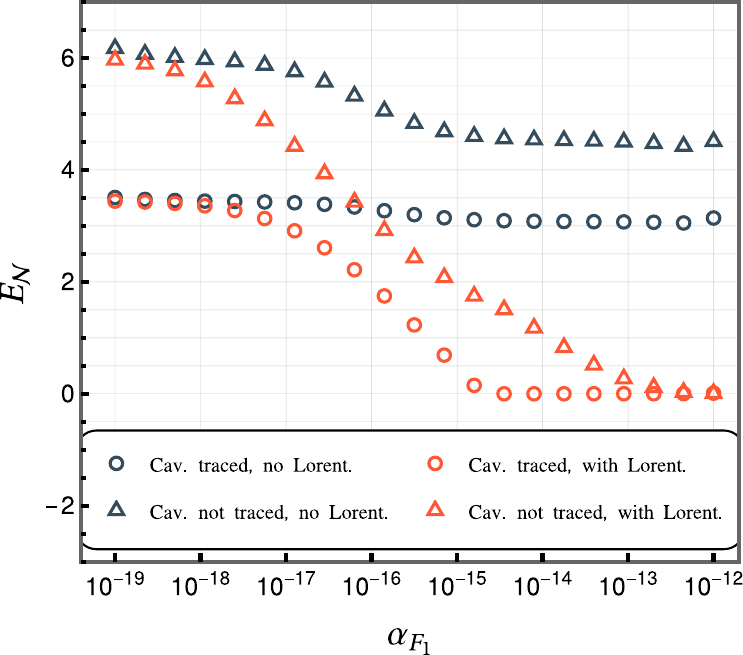}
\caption{The effect of resonant modes on the logarithmic negativity, $E_\mathcal{N}$, for both partitions. We see that the negativity reduces with increasing $\alpha_{F_1}$ or with the introduction of resonant modes. Note that $\alpha_{F_2} = 1$. \label{fig:violin_effect} }
\end{figure}

We investigate the effect of the resonant modes on the entanglement. In Fig. \ref{fig:violin_effect}, with low force noise guaranteeing well-behaved numerics and setting $\alpha_{F_2} = 1$, we plot $E_{\mathcal{N}}$ for noise curves with and without these modes (orange and gray, resp.) and for both partitions (circles when the cavity is traced out). We see explicitly here that there is more entanglement in the partition where we do not trace over the cavity. For low noise (low $\alpha_{F_1}$), the negativity remains unchanged, hence the resonant modes do not affect the entanglement significantly in this regime. As the level of noise increases, resonant modes cause the logarithmic negativity to decrease faster than the negativities calculated without the resonant modes for both partitions. The system becomes separable when resonant modes are included for $\alpha_{F_1} \approx 10^{-15}$ and $\alpha_{F_1} \approx 10^{-12}$ when we do and do not trace over the cavity, respectively. It seems reasonable to expect that entanglement will not emerge when force noise becomes stronger. Hence, extrapolating $\alpha_{F_1} \to 1$, this is evidence (but not a rigorous proof) that aLIGO in its current operation regime (but without squeezed input) probably contains no optomechanical entanglement.
\vspace{-1ex}
\section{Discussion and Conclusions}
\vspace{-2ex}
In this paper, we developed a framework to determine and quantify bipartite entanglement in an optomechanical system in the non-sideband-resolved regime, in the presence of non-Markovian Gaussian noises. The main novelty of our work is to enable the study of non-Markovian noise drives, which are common in devices with low mechanical frequencies, typically associated with large/macroscopic masses. Hence, we focused on macroscopic entanglement and used a free mass with structural damping and coating Brownian noise as an initial example, and then aLIGO as a more detailed case study. However, non-Markovian noise can also be found in optomechanical systems with higher frequencies \cite{Evidence_for_structural_damping, non_markovian_micromech_brownian, Gröblacher2009, meng_measurement_2022}, and our framework can be applied to them in the same way. Besides improving the quality and accuracy of predictions, modeling non-Markovian dynamics is rich in interesting physics and possibly useful phenomena: for example, in Ref.~\cite{meng_measurement_2022}, they find that squashing/squeezing the mechanical state is less demanding in the presence of structural damping compared to Markovian viscous damping.

We tested for bipartite entanglement by looking at the separability between the mechanical oscillator and (i) the outgoing light field, and (ii) the joint system of the cavity and the outgoing light field. For low levels of classical noise, we saw that the latter partition is more entangled compared to the former. However, for high levels of classical noise, we did not see a significant advantage in using one partition over the other.

In the low mechanical frequency regime, where the free-mass limit approximation holds, we found that the presence of entanglement is independent of the coherent optomechanical interaction strength and depends only on the relative strength of force and sensing noise. This result is similar to that already found in \cite{Miao2010} for white noise drive. Beyond the free-mass limit approximation, by parametrizing the noise curves of aLIGO, we were able to find a region of noise curves where entanglement exists, and we showed that there is a trade-off between the overall noise level and the cutoff frequency. Due to the high level of the current classical noise in aLIGO, we were not able to reach a definite conclusion in terms of the existence of entanglement in the system, even though it's unlikely to have a significant amount of entanglement based on our simulations. However, we saw that entanglement exists if we assume a system without seismic noise, even when the suspension thermal noise is still present. This is an important result since it shows that classical noises, even at very low frequencies, are able to demolish entanglement. 

We also looked at how resonances in the noise curves of the system affect the amount of entanglement, and we saw that entanglement is more resistant (i.e. it disappears for higher levels of classical noise) for the partition where we test for the separability between the mechanical oscillator and the joint system of the cavity and the outgoing light field. For future work, we plan to develop better sampling strategies to overcome numerical instabilities.
\vspace{-1ex}
\begin{acknowledgments}
\vspace{-2ex}
We thank the Chen Quantum Group for helpful discussions. S.D. and Y.C. acknowledge the support by the Simons Foundation (Award No. 568762). K.W., C.G., and M.A. received funding from the European Research Council (ERC) under the European Union’s Horizon 2020 research and innovation program (Grant Agreement No 951234), and from the Research Network Quantum Aspects of Spacetime (TURIS). K.H. was supported by the Deutsche Forschungsgemeinschaft
(DFG, German Research Foundation)
through Project-ID 274200144 SFB 1227 (projects A06) and Project-ID 390837967-EXC 2123.

\end{acknowledgments}


\appendix

\renewcommand{\arraystretch}{1.}
\vspace{-2ex}
\section{Transformation of the Optomechanical Hamiltonian}
\label{app:travelling_wave}
\vspace{-2ex}
Strictly speaking, the Hamiltonian in Eq. (\ref{eq:first_hamiltonian}) is written in terms of Schr\"odinger operators.  The symbol $\Omega$ in Eq.~\eqref{eq:first_hamiltonian} is used to label a spatial mode which has a reduced wavenumber of $\Omega/c$, and a free oscillation frequency of $\Omega$.  
More specifically, $\omega_0+\Omega$ is used to indicate a spatial mode whose wave number is $(\omega_0+\Omega)/c$, where $c$ is the speed of light.  We can also write the same Hamiltonian in the spatial domain. Following \cite{Chen_2013} and setting $c=1$, we define
\begin{equation}
\label{Eq:spatial_light}
\begin{gathered}
\hat{c}(z) = \int_{-\infty}^{\infty} \frac{d \Omega}{2 \pi} \hat{c}(\omega_0+\Omega) e^{i \Omega z}  \,,
\end{gathered}
\end{equation}
which represents a spatial mode of the light field with wave number $\Omega$ --- or a temporal mode of the free light field with frequency $\Omega$, since we assume no dispersion.  The Hamiltonian can then be rewritten  as
\begin{equation}
\begin{gathered}
H = \hbar \omega_m \hat{B}^\dagger \hat{B} + \hbar \Delta \hat{A}^\dagger \hat{A} - \hbar G \hat{x} (\hat{A}^\dagger + \hat{A}) \\ + i \hbar \sqrt{2 \gamma}
     [ \hat{A}^\dagger \hat{c}(z=0) - \hat{A} \hat{c}^\dagger(z=0) ] \\ - i \hbar \int_{- \infty}^{\infty} \hat{c}^\dagger(z) \partial_z \hat{c}(z) \, d z \,.
\end{gathered}
\end{equation}

The only non-zero commutators for the creation and annihilation operators, as well as the spatial modes of the light field are
\begin{subequations}
\begin{align}
\relax [\hat{c}(\omega_0+\Omega), & \hat{c}^\dagger(\omega_0+\Omega')] =  2 \pi \delta(\Omega-\Omega') \,, \\
[ \hat{c}(z), & \hat{c}^\dagger(z') ] = \delta(z-z') \,,
\end{align}
\end{subequations}
where $\delta$ is the Dirac delta distribution.
\vspace{-2ex}
\section{Parameter Tables}
\label{appendix:parameter_tables}
\vspace{-2ex}
This appendix contains the tables with the numerical values of the parameters used in modeling the classical and quantum noise curves, including the resonant modes. The parameters were calculated by minimizing the mean-squared error between the actual noise curves of aLIGO taken from pygwinc and the theoretical models, characterized by the parameters of interest, sampled logarithmically in frequency. Table \ref{tab:noiseParams} contains the parameters for the classical force and sensing noise in aLIGO defined in Eqs. (\ref{Eq:S^{LIGO}_{F}_s_x}), Table \ref{tab:ligoParams} contains the parameters of aLIGO introduced in the optomechanical Hamiltonian of Sec. \ref{sec:system_dynamics}, and lastly, Table \ref{tab:violinModeParams} contains the parameters of the resonant modes present in aLIGO's noise curves, modeled with Eq. (\ref{Eq:S^{LIGO}_{F}_lorentzians}).

\begin{table}
\caption{Classical Noise Model Parameters}
\label{tab:noiseParams}
\begin{tabular}{lccc}
\hline \hline
 Parameter & Symbol & Value & Units
\\ \hline
\rule{0pt}{3ex}
Force Noise Time Constant & $ \tau_F $ & $ 1.6 \cdot 10^{-20}$ & s \\
Force Noise Cutoff Frequency & $ \omega_F $ & $2\pi \cdot 0.25 $ & rad/s \\
Sensing Noise Time Constant 1 & $ \tau_{X_1} $ & $10^{-50}$ & s \\
Sensing Noise Time Constant 2 & $ \tau_{X_2} $ & $10^{-48}$ & s \\
Sensing Noise Cutoff Frequency & $ \omega_X $ & $2\pi \cdot 10^{4}$ & rad/s \\
\makecell[l]{ Suspension Thermal Noise \\ Time Constant} & $ \tau_{ST} $ & $ 3.1 \cdot 10^{-35}$ & s \\
\makecell[l]{Suspension Thermal Noise \\ Cutoff Frequency} & $ \omega_{ST} $ & $2\pi \cdot 1.9 \cdot 10^3 $ & rad/s \\
\hline \hline
\end{tabular}
\end{table}
\vspace{-2ex}
\section{Noise Model}
\label{appendix:noise_model}
\begin{figure}
\includegraphics[width=0.85 \columnwidth]{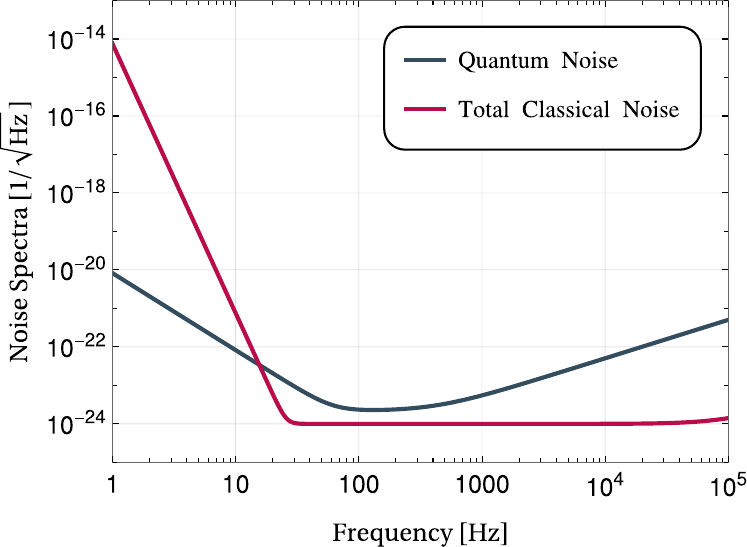}
\caption{\label{fig:total_noise_spectrum} Modeled noise spectrum for aLIGO.}
\end{figure}

\vspace{-2ex}
Our model for the total classical and quantum noise in aLIGO is displayed in Fig. \ref{fig:total_noise_spectrum}. Note that the coating Brownian noise, which is the dominating classical noise source above $\sim 10$ Hz, is modeled in a piecewise manner: a white noise in the $10-10^5$ Hz band and a noise source increasing as $\Omega^2$ for frequencies larger than $10^5$ Hz. Even though the sensing noise in the $10-10^5$ Hz band decreases as $\Omega^{0.8}$ in the power spectral density, we choose to model it with a frequency-independent white noise for simplified calculations. Our choice is justified since the quantum noise dominates over the coating Brownian noise in that region.

\begin{table}
\caption{aLIGO Parameters}
\label{tab:ligoParams}
\begin{tabular}{lccc}
\hline\hline
Parameter & Symbol & Value & Units
\\ \hline
\rule{0pt}{3ex}
Mechanical Resonant Frequency & $\omega_m$ & 2$\pi \cdot $ 0.9991 & rad/s \\
Mirror Mass & $M$ & 9.446 & kg\\
Cavity Decay Rate & $\gamma$ & 2$\pi \cdot $ 424.6 & rad/s\\
Arm Length & $L$ & 3.995 & km \\
Circulating Power & $P_c$ & 322.7 & kW \\
Laser Wavelength & $ \lambda $ & 1064 & nm \\
Mechanical Damping & $ \gamma_m $ & 2$\pi \cdot 10^{-3}$ & rad/s \\
\hline\hline
\end{tabular}
\end{table}

\begin{table}
\caption{Resonant Mode Parameters}
\label{tab:violinModeParams}
\begin{tabular}{lccc}
\hline\hline
 Parameter & Symbol & Value & Units
\\ \hline
\multirow{7}{*}{Mode Frequency} & $\Omega_{v_1}$ & 2$\pi \cdot 0.441$ & rad/s \\
& $\Omega_{v_2}$ & 2$\pi \cdot 0.995$ & rad/s \\
& $\Omega_{v_3}$ & 2$\pi \cdot 1.98$ & rad/s \\
& $\Omega_{v_4}$ & 2$\pi \cdot 2.37$ & rad/s \\
& $\Omega_{v_5}$ & 2$\pi \cdot 3.38$ & rad/s \\
& $\Omega_{v_6}$ & 2$\pi \cdot 3.81$ & rad/s \\
& $\Omega_{v_7}$ & 2$\pi \cdot 9.73$ & rad/s \\
[-1.8ex] \\ \hline \\ [-1.8ex]
\multirow{7}{*}{Full Width at Half Maximum} & $\Gamma{v_1}$ & 2$\pi \cdot 1.92 \cdot 10^{-3}$ & rad/s \\
& $\Gamma{v_2}$ & 2$\pi \cdot 5.63 \cdot 10^{-5}$ & rad/s \\
& $\Gamma{v_3}$ & 2$\pi \cdot 2.11 \cdot 10^{-5}$ & rad/s \\
& $\Gamma{v_4}$ & 2$\pi \cdot 1.44 \cdot 10^{-1}$ & rad/s \\
& $\Gamma{v_5}$ & 2$\pi \cdot 1.45 \cdot 10^{-4}$ & rad/s \\
& $\Gamma{v_6}$ & 2$\pi \cdot 1.65 \cdot 10^{-3}$ & rad/s \\
& $\Gamma{v_7}$ & 2$\pi \cdot 1.03 \cdot 10^{-3}$ & rad/s \\ \hline \\
\multirow{7}{*}{Amplitude} & $A_{v_1}$ & 159 & rad/s \\
& $A_{v_2}$ & 93.8 & rad/s \\
& $A_{v_3}$ & 538 & rad/s \\
& $A_{v_4}$ & 235 & rad/s \\
& $A_{v_5}$ & 353 & rad/s \\
& $A_{v_6}$ & 27.4 & rad/s \\
& $A_{v_7}$ & 78.0 & rad/s \\
\hline\hline
\end{tabular}
\end{table}

\vspace{-2ex}
\section{Structure of the Covariance Matrix}
\label{app:cov_matrix}
\vspace{-2ex}
The mirror and the cavity, at $t=0$, constitute two modes, whereas there are an infinite number of modes in the outgoing light field given by the quadratures $\hat{v}_{1}(t),\hat{v}_{2}(t)$, $t \in (-\infty, 0)$. In continuum coordinates, we can first write down the commutator matrix,
\begin{equation}
\mathbf{K} = \left[
\begin{array}{ccc}
\mathbf{K}^B \\
& \mathbf{K}^A \\
& & \mathbf{K}^v
\end{array}
\right] \,,
\end{equation}
with
\begin{align}
    \mathbf{K}^A &=\mathbf{K}^B  =
    \left[\begin{array}{cc}
    0 & i \\
    -i & 0
    \end{array}\right]\,,\nonumber\\
    \mathbf{K}^v &=
     \left[\begin{array}{cc}
    0 & i\delta(t-t') \\
    -i\delta(t-t') & 0
    \end{array}\right] \,.
\end{align}
Note that $\mathbf{K}^v$ is a $2\times 2$ block matrix, but each block is infinite dimensional, with columns and rows indexed by $t$ and $t'$, respectively.  The indices $t$ and $t'$ each run through all negative real numbers,  $(-\infty, 0]$.  In order to represent these covariance matrices in a less ambiguous and more operational way, we shall adopt an index notation, in which $j,k,l,m$ are discrete and run through 1 and 2, while $t$ and $t'$ are continuous and run through negative real numbers.  We can then write  
\begin{equation}
    K^A_{jk} =i\epsilon_{jk}\,,\; K^B_{lm}=i \epsilon_{lm}\,,\;
K^v_{lt,mt'}=i\epsilon_{lm}\delta(t-t') \,.   
\end{equation}
Note that for $\mathbf{K}^v$ we have a two-dimensional row index $(l,t)$, as well as a two-dimensional column index $(m,t')$, to label quadrature and arrival time. 

The covariance matrix $\mathbf{V}$ of the system contains the steady state correlations between the mechanical mode at $t=0$, the cavity mode at $t=0$, and the outgoing light field modes for $t<0$. It can be written as 
\begin{equation}
\label{cov_matrix}
\mathbf{V} =  \left[ 
\begin{array}{cc|c}
V^{BB} & V^{BA} & V^{Bv} \\
V^{AB} & V^{AA} & V^{Av} \\
\hline
V^{vB}& V^{vA} & V^{vv}
\end{array}
\right] \equiv 
\left[
\begin{array}{cc}
V^{QQ}  & V^{Qv} \\
V^{vQ} & V^{vv}
\end{array}
\right] \,.
\end{equation}
Here each of $A$ and $B$ represent two dimensions, while $v$ represents an infinite number of dimensions.  For computational purposes we can group $A$ and $B$ together as $Q$, or 
\begin{equation}
\hat Q_J = \hat Q_{(1,2,3,4)} = (\hat B_1, \hat B_2,\hat A_1,\hat A_2) \,.
\end{equation}
Here we shall use upper-case Latin indices to run from 1 to 4, to label quadratures in the mechanical and cavity modes.  We can write
\begin{equation}
V^{QQ}_{JK} = \frac{1}{2} \langle \hat Q_J \hat Q_K + \hat Q_K \hat Q_J \rangle = \int_{0}^{\infty}\frac{d\Omega}{2\pi}S_{Q_J Q_K}(\Omega) \,,
\end{equation}
where the (one-sided) cross spectrum is defined via
\begin{equation}
S_{XY}(\Omega)\delta(\Omega-\Omega') = 2 \pi \langle \hat X(\Omega) \hat Y^\dagger(\Omega') + \hat Y^\dagger(\Omega') \hat X(\Omega) \rangle \,, 
\end{equation}
and they can be obtained from solutions to the Heisenberg Equations, as well as the spectra 
\begin{equation}
\label{Eq:vacuum_spectra}
    S_{u_i u_j} (\Omega)=\delta_{ij} \,,
\end{equation}
and the prescriptions we use for the spectra of $S_{n_X}(\Omega)$ and $S_{n_F}(\Omega)$. The uncorrelated white spectra between $\hat u_1$ and $\hat u_2$ in Eq. (\ref{Eq:vacuum_spectra}) result from the ingoing light field being in its vacuum state. We shall discuss the magnitude and frequency dependence of $S_{n_X}(\Omega)$ and $S_{n_F}(\Omega)$ in depth in the next appendix.

For elements that involve $v$, we shall still use $lt$ for column indices and $mt'$ for row indices.  We then have
\begin{align}
    V^{Qv}_{J,mt'} &= \frac{1}{2} \langle \hat Q_J \hat v_{m}(t') + \hat v_{m}(t') \hat Q_J \rangle \nonumber \\ &= \frac{1}{2} \int\limits_{-\infty}^{+\infty} \frac{d\Omega}{2\pi} S_{Q_J v_m}(\Omega) e^{i\Omega t'} \,,
\end{align}
and
\begin{align}
    V^{vv}_{lt,mt'} &= \frac{1}{2} \langle \hat v_l(t) \hat v_{m}(t') + \hat v_m(t) \hat v_{l}(t') \rangle \nonumber \\ &= \frac{1}{2} \int\limits_{-\infty}^{+\infty} \frac{d\Omega}{2\pi} S_{v_l v_m}(\Omega) e^{-i\Omega(t- t')} \,. 
\end{align}
Note that $V^{Qv}_{J,mt'} = V^{vQ}_{mt',J}$.

In numerical computations, we will have to convert the continuum of $t,t'\le 0$ into a finite grid. This means we will sample a finite duration $T$ with a step size of $\Delta t$. We shall still use lower-case Latin indices to run from 1 to 2, and upper case Latin indices to run from 1 to 4, while we use Greek indices, for example, $\alpha = 0,1,2,\ldots, T/\Delta t \equiv \mathcal{N}-1$, to replace $t$. We shall write
\begin{equation}
\label{eqKdiscrete}
    K_{lt,mt'}^{v} \rightarrow K^v_{l\alpha,m\alpha'} = i \epsilon_{lm}\delta_{\alpha\alpha'} \,.
\end{equation}
Note that a Kronecker delta now replaces the Dirac delta.  For the covariance matrix, we replace
\begin{align}
\label{Eq:VQvdiscrete}
V_{J,mt'}^{Qv} \rightarrow V_{J,m\alpha'}^{Qv} &= \frac{\sqrt{\Delta t} }{2} 
\langle \hat Q_J \hat v_m(t_{\alpha'}) + \hat v_m(t_{\alpha'})  \hat Q_J \rangle \,,
\end{align}
which are $4\times 2\mathcal{N}$  and $2\mathcal{N}\times 4$ dimensional matrices (with $V_{J,mt'}^{vQ}$), and  
\begin{align}
\label{Eq:eqVvvdiscrete}
V_{lt,mt'}^{vv} \rightarrow V_{l\alpha,m\alpha'}^{vv} &= \frac{\Delta t}{2} 
\langle \hat v_l(t_\alpha) \hat v_m(t_{\alpha'}) + \hat v_m(t_\alpha) \hat v_l(t_{\alpha'}) \rangle \,,
\end{align}
which is a $2\mathcal{N} \times 2\mathcal{N}$-dimensional matrix.  For the discrete sampling times we have defined
\begin{equation}
t_{\alpha} =- \left( \alpha+\frac{1}{2}\right)\Delta t\,,
\end{equation}
where the additional $\frac{1}{2} \Delta t $ provides a faster convergence in numerics. The entire covariance matrix is then $(2\mathcal{N}+4)\times (2\mathcal{N}+4)$ - dimensional. 

Our particular convention of inserting $\Delta t$ at various places of the matrix is associated with our convention of discretizing vectors.  For a generic variable, in the continuum form, we can always express it as 
\begin{align}
    X &= \alpha^j \hat A_j + \beta^j \hat B_j + \int_{-\infty}^0 \xi^j(t) \hat v_j(t) dt \nonumber\\
    &= \gamma^J \hat Q_J +  \int_{-\infty}^0 \xi^j(t) \hat v_j(t) dt \,,
\end{align}
where we have used upper indices for vector components, and we have grouped $\alpha^j$ and $\beta^j$ into $\gamma^J$.  The variance of $X$, which is formally written as $\mathbf{X}^\dagger \mathbf{V}\mathbf{X}$, will then be
\begin{align}
\label{eq:XVX1}
\mathbf{X}^\dagger \mathbf{V}\mathbf{X} &= \frac{1}{2} \gamma^J \langle \hat Q_J \hat Q_K + \hat Q_K \hat Q_J \rangle \gamma^K \nonumber\\ &+ \int_{-\infty}^0 \gamma^J \langle \hat Q_J \hat v_m(t') + \hat v_m(t') \hat Q_J \rangle\xi^m(t') dt' \nonumber\\
&+ \frac{1}{2} \iint_{-\infty}^0 \xi^l(t) \langle \hat v_l(t) \hat v_m(t') + \hat v_m(t) \hat v_l(t') \rangle \xi^m(t')  dt dt' \,.
\end{align}

As we convert the integrals in Eq.~\eqref{eq:XVX1} into summations, $\int$ will become $\Sigma$, while $dt$ will become $\Delta t$. We shall take
\begin{equation}
    \xi^{m\alpha} =\xi^m(t_\alpha)\sqrt{\Delta t} .
\end{equation}

Together with Eq. (\ref{Eq:VQvdiscrete})--(\ref{Eq:eqVvvdiscrete}), the fully discretized version of Eq. (\ref{eq:XVX1}) will then be

\begin{align}
\label{eq:XVX2}
\mathbf{X}^\dagger \mathbf{V}\mathbf{X} = \gamma^J V^{QQ}_{JK} \gamma^K + \gamma^J V^{Qv}_{J,m \alpha}\xi^{m\alpha} &+ \xi^{l\beta} V^{vQ}_{l\beta,K}\gamma^K \nonumber \\ &+ \xi^{l\alpha} V^{vv}_{l\alpha,m\beta} \xi^{m\beta}.
\end{align}
In this convention, the usual vector norm for the discretized version of a function of time coincides with the $L^2$-norm of that function.  It can also be checked that discretized matrices in Eqs.~\eqref{eqKdiscrete}--\eqref{Eq:eqVvvdiscrete}, when contracted with vectors in this convention, lead to the appropriate integrals. Note that if a $\delta(t_\alpha-t_\alpha')$ shows up in Eq.~\eqref{Eq:eqVvvdiscrete}, we will take $\Delta t \delta(t_\alpha-t_\alpha')\rightarrow \delta_{\alpha\alpha'}$, as in Eq.~\eqref{eqKdiscrete}.

Corresponding to the discussion at the end of Sec.~\ref{sec:entanglement_criteria_partitions} (also shown in Fig. \ref{fig:partitions}), here we consider entanglement between: (i) mass at $t=0$ and the out-going light field that had emerged during $t\le 0$  and (ii) mass and the joint system of the cavity mode as well as light that had emerged during $t\le 0$.  In case (i), we simply throw away elements involving $A$ in both $\mathbf{K}$ and $\mathbf{V}$, while in case (ii) we consider the full matrices. In both cases, $\mathbf{V}_{\rm pt}$ is obtained by adding a minus sign to the column involving $\hat B_2$ and the row involving $\hat B_2$ -- but not the diagonal element at which they intersect.

\vspace{-2ex}
\section{Numerical Implementation for aLIGO's Noise}
\label{sec:numerical_implementation}
\vspace{-2ex}
In our simulations, we use $dt = 0.25 $ ms and $T = 0.1 $ s, which corresponds to sampling the light field at 4000 Hz and working with the outgoing field emitted from the cavity between $t = -0.1 $ s and $t = 0 $ s. We achieve numerical convergence with these parameters. To quantify the amount of entanglement in the system, we use the logarithmic negativity defined in Eq. (\ref{Eq:negativity}). However, this is possible only for low levels of classical noise. For high levels of classical noise, classical correlations dominate over quantum correlations, which causes the cross-correlation values in the system to cover a wide-range of orders of magnitude, mostly due to the 14\textit{th} power of $\Omega$ in our force noise model Eq. (\ref{Eq:S^{LIGO}_{F}_s_x}). For aLIGO parameters, the entries of the covariance matrix span about 20 orders of magnitudes, while we attempt to find a symplectic eigenvalue of order 1 -- this is numerically an extremely challenging problem.

\begin{figure}
\includegraphics[width=0.8\columnwidth]{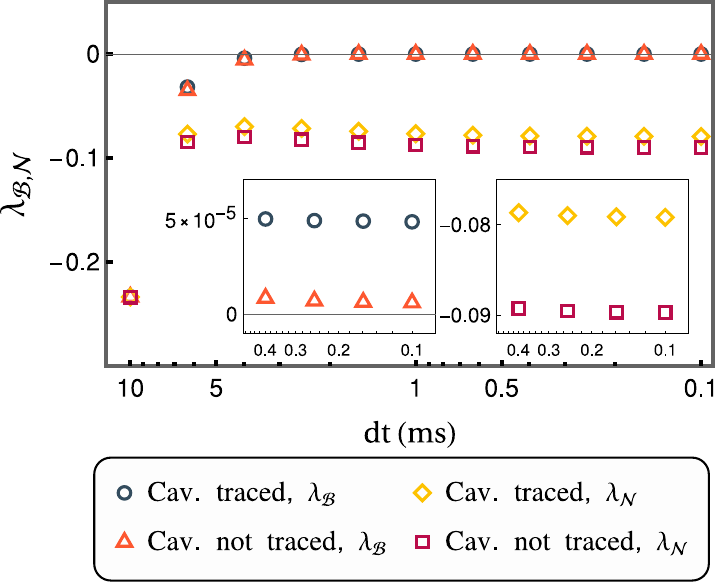} \\
\vspace{0.5cm}
\includegraphics[width=0.8\columnwidth]{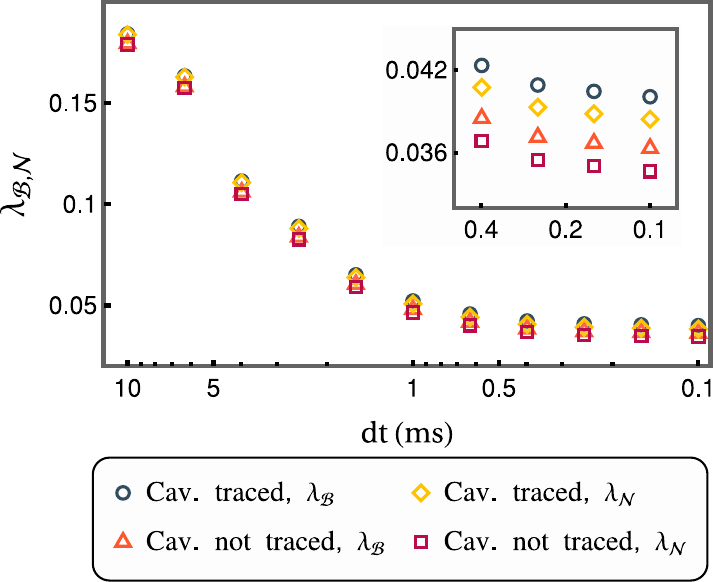}
\caption{\label{fig:lambdas_vs_dt_all} $\lambda_{\mathcal{B}}$ and $\lambda_{\mathcal{N}}$ for $\alpha_{F_1} = 10^{-15}$ and $\alpha_{F_2} = 15 $ [top row, (a)]; $\alpha_{F_1} = 10^{-8}$, $\alpha_{F_2} = 10 $, and $\alpha_{X_2} = 10^{3}$ [bottom row, (b)]. We set $T= 0.1$ s, and $dt \in [0.1, 10]$ ms. Note that $dt$ decreases to the right of the plots. In Figure a, entanglement exists for both partitions, and $\lambda_{\mathcal{B}}$($\lambda_{\mathcal{N}}$) is re-plotted in the inset on the left(right) for $dt \in [0.1, 0.5]$ ms. In Figure b, the state is separable for both partitions, and $\lambda_{\mathcal{B}}$ and $\lambda_{\mathcal{N}}$ are re-plotted in the inset for $dt \in [0.1, 0.5]$ ms.}
\end{figure}

Numerical errors also arise because of time-binning with an insufficient resolution. Thus, we lose precision on the numerically determined covariance matrices. This affects the smallest symplectic eigenvalue $\tilde{\nu}_{min}$ to the point that it cannot be used to measure entanglement with the logarithmic negativity. Numerical imprecision can lead to covariance matrices that do not satisfy Heisenberg uncertainty bound Eq.~\eqref{Eq:Heisenberg_uncertainty}, they thus do not correspond to a bona fide state and we call them \emph{non-physical}. One way to get around this loss of precision is to use the PPT criterion as a yes/no test only, renouncing the magnitude information of $\tilde{\nu}_{min}$. The sampling frequency during time binning should be higher than the Nyquist rate of the system (i.e. twice the largest frequency in the system), since the entries of the covariance matrix contain correlations from all frequencies. In our system, the largest frequency is the cavity decay rate $\gamma=424$ Hz. Therefore, we choose $dt<1/(2\cdot 424) \approx 1.2$ ms. However, as we decrease $dt$, we are limited by computational resources, such as the RAM size, or time. The parameter $dt$ is limited to an optimal range determined by this trade-off. We thus develop the following strategy: we first quantify the amount of numerical errors in the system by computing the most negative eigenvalue (if it exists) of $\mathbf{V} + \frac{1}{2} \mathbf{K}$ before and after the partial transpose operation, denoted as $\lambda_{\mathcal{B}}$ and $\lambda_{\mathcal{N}}$, respectively. Then, we decide that entanglement exists if $\lambda_{\mathcal{B}} > 0$ and $\lambda_{\mathcal{N}} < 0$, or if $\lambda_{\mathcal{B}} < 0$ and $ |\lambda_{\mathcal{N}}| \gg |\lambda_{\mathcal{B}}| $. Furthermore, we decide that the system is separable if $\lambda_{\mathcal{B}} > 0$ and $\lambda_{\mathcal{N}} > 0 $. 

Two case studies about this strategy can be found in Fig. \ref{fig:lambdas_vs_dt_all} where we fix $T = 0.1 $ s, change $dt$, and examine how $\lambda_{\mathcal{N}}$ and $\lambda_{\mathcal{B}}$ change by plotting $\lambda_{\mathcal{N}}$ and $\lambda_{\mathcal{B}}$. We decide on entanglement if $\lambda_{\mathcal{B}} \geq 0$ and $\lambda_{\mathcal{N}} < 0$, or $\lambda_{\mathcal{B}} < 0$, $\lambda_{\mathcal{N}} < 0$, and $ |\lambda_{\mathcal{N}}| \geq 100 |\lambda_{\mathcal{B}}|$. Our criteria for convergence is a relative change smaller than $5 \%$ for both $\lambda_{\mathcal{N}}$ and $\lambda_{\mathcal{B}}$ as we change $dt$. In Fig. \ref{fig:lambdas_vs_dt_all}a, we work with a low level of classical force noise and set $\alpha_{F_1} = 10^{-15}$, $\alpha_{F_2} = 15 $ in Eq.~\eqref{Eq:S^{LIGO}_{F}_s_x}. We see that $\lambda_{\mathcal{N}}$ and $\lambda_{\mathcal{B}}$ converge with $\lambda_{\mathcal{N}}$ changing by $0.053 \%$, $0.068 \%$, and $\lambda_{\mathcal{B}}$ changing by $4.9 \%$, $0.77 \%$ before and after tracing over the cavity, respectively, for $dt = 0.1$ ms. The system is entangled for both partitions since $\lambda_{\mathcal{B}} \geq 0$ and $\lambda_{\mathcal{N}} < 0$. Furthermore, $\lambda_{\mathcal{B}}$ becomes positive and converges after $dt \sim 1$ ms, or a sampling frequency of 1000 Hz; consistent with the discussion above relating physicality to Nyquist rate of $\sim 850$ Hz. We also see that $\lambda_{\mathcal{N}}$ converges for similar values of $dt$ from Fig. \ref{fig:lambdas_vs_dt_all}a.

\begin{figure}
    \centering
    \includegraphics[width=0.9\columnwidth]{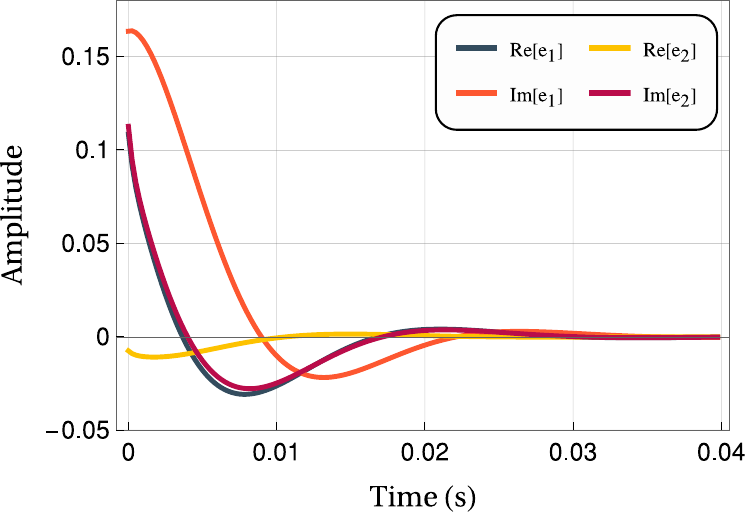}
    \caption{The real and imaginary parts of the eigenvector for the negative eigenvalue of $\mathbf{V}_{\rm pt} + \frac{1}{2} \mathbf{K}$ for $\alpha_{F_1} = 10^{-15}$, $\alpha_{F_2} = 15 $, denoted by $e_1$ and $e_2$, which correspond to the first and second halves of the entire eigenvector, respectively. The reason behind this slicing is the block-matrix structure of the light-field sector of the covariance matrix. Furthermore, since the light field modes are continuous in time, $e_1$ and $e_2$ are also functions of time.}
    \label{fig:eigenvector}
\end{figure}

In Fig. \ref{fig:eigenvector}, we plot the light-field section of the eigenvector associated with the (converged) minimal eigenvalue of $\mathbf{V}_{\rm pt} + \frac{1}{2} \mathbf{K}$, for the partition where we do not trace over the cavity and with $\alpha_{F_1} = 10^{-15}$, $\alpha_{F_2} = 15$. It corresponds to a temporal mode of the free electromagnetic field outside the cavity. It is that particular mode associated with the (sole) negative eigenvalue that is entangled with the joint system mechanics plus cavity. The four curves correspond to the real and imaginary parts of $\hat{v}_1(t)$ and $\hat{v}_2(t)$. They have the form of smooth decaying oscillations with the same frequency and decay rate, but differing by a phase. This form of the mode functions was predicted for a white force noise in \cite{universal_quantum_entanglement}; which gives us confidence in the correctness of our study. Also, exponentially decaying demodulation pulses were used to demonstrate optomechanical entanglement \cite{Mechanical_Motion_with_Microwave_Fields, Quantum_entanglement_teleportation_pulsed} and proposed for a demonstration in the stationary regime \cite{corentin_klemens_stat_optomech_entanglement}. We fit functions in the form of $e^{-\gamma_* t}\sin{(\omega_* t + \theta)}$ to each curve, which results in $\omega_*/(2\pi)  \approx 40 $ Hz, and $\gamma_*/(2\pi) \approx 25 $ Hz. In the frequency domain, exponentially decaying harmonic oscillations are Lorentzians, centered at $\pm \omega_*$ and with a bandwidth (FWHM) $2 \gamma_*$. In aLIGO's noise budget (Fig.~\ref{fig:aLIGO_noise_curves}), these Lorentzians are on the low frequency side of the low-noise band and their halfwidth at half-maximum to the left crosses the quantum noise, where it is not yet dominated by suspension thermal and seismic noises---although we saw in Sec.~\ref{sec:effect_of_force_noise} that the latter is probably the main mechanism preventing optomechanical entanglement. We add that Lorentzians are heavy-tail distributions, being a possible reason why even lower frequency components matter.

\begin{figure}
\includegraphics[width=0.6\columnwidth]{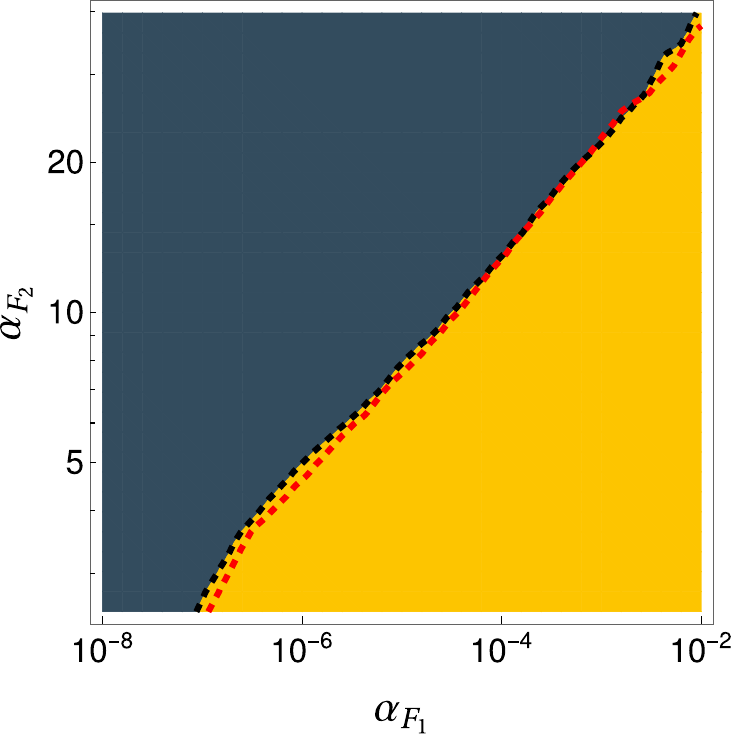}
\caption{\label{fig:contour_plots} Contour plot depicting the effect of the force noise spectrum on our numerical precision for both partitions. The force noise levels increase toward the bottom-right of the figure. The black and the red dashed lines separate the regions where numerics converge from the regions where numerics fail for the partition where we do and do not trace over the cavity, respectively. The region where numerics converge for both partitions is marked in gray, whereas numerics fail for both partitions toward the bottom right of the figure, past the red dashed line, in the yellow region. 
}
\end{figure}

\begin{figure}
\includegraphics[width=0.8\columnwidth]{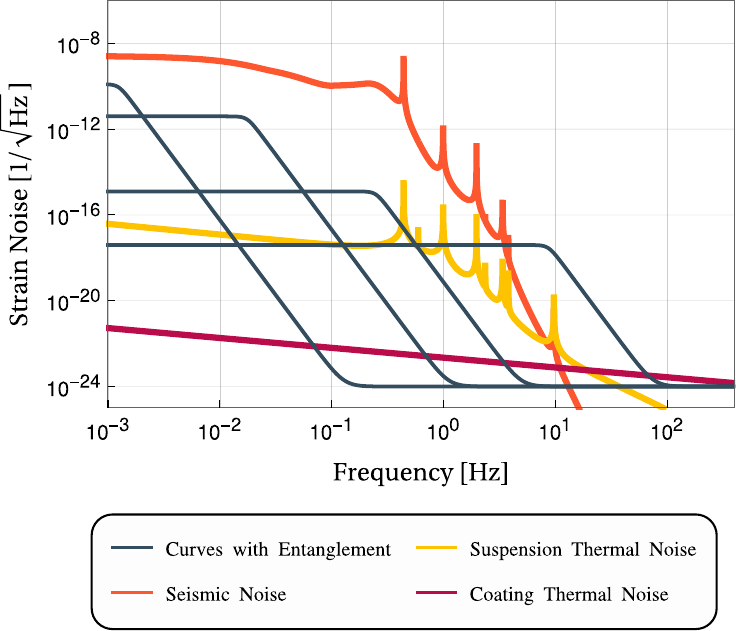}
\caption{Examples of force noise spectra along the boundary where we lose numerical precision (defined with $ |\lambda_{\mathcal{N}}| = 100 |\lambda_{\mathcal{B}}|$) for the partition where we do not trace over the cavity, plotted in black. aLIGO noise curves are also plotted for comparison. The black curves are parametrized with (from left to right) $\alpha_{F_1} = 1$ and $\alpha_{F_2} = 200$, $\alpha_{F_1} = 10^{-3}$ and $\alpha_{F_2} = 14$, $\alpha_{F_1} = 10^{-10}$ and $\alpha_{F_2} = 1$, and $\alpha_{F_1} = 10^{-15}$ and $\alpha_{F_2} = 0.029$. \label{fig:noise_curves_with_entanglement} }
\end{figure}

In Fig. \ref{fig:lambdas_vs_dt_all}b, we set $\alpha_{F_1} = 10^{-8}$, $\alpha_{F_2} = 10 $, and $\alpha_{X_2} = 10^{3}$, causing the sensing noise to dominate over quantum noise for frequencies in the $30-2000$ Hz band. We again see that $\lambda_{\mathcal{N}}$ and $\lambda_{\mathcal{B}}$ converge with $\lambda_{\mathcal{N}}$ changing by $1.3 \%$, $1.2 \%$, and $\lambda_{\mathcal{B}}$ changing by $1.2 \%$, $1.1 \%$ before and after tracing over the cavity, respectively, for $dt = 0.1$ ms. Since $\lambda_{\mathcal{B}}$ and $\lambda_{\mathcal{N}}$ are positive for both partitions, we conclude that there is no entanglement in the system for either partition. When we increase the classical noise level we see that convergence is much harder to achieve. Furthermore, $\lambda_{\mathcal{B}}$ and $\lambda_{\mathcal{N}}$ are negative, and $\lambda_{\mathcal{B}} \sim \lambda_{\mathcal{N}}$ for every value of $dt$ regardless of when we do or do not trace over the cavity. Therefore, we cannot conclude that there is entanglement for either of the partitions. These case studies show that we can use $\lambda_{\mathcal{B}}$ as an indicator of the ``non-physicality" of the covariance matrix $\mathbf{V}$ introduced by finite-sampling and high levels of classical noise in the system, and that the negativity of $\lambda_{\mathcal{N}}$ is not enough to decide on entanglement when we consider the numerics.

For our model of aLIGO's non-Markovian noises, Eqs.~\eqref{Eq:S^{LIGO}_{F}_s_x}, we study the numerical stability of broad noise regimes, parametrized by the pair $\alpha_{F_j}$, $j=1,2$. We set $\alpha_{X_j} = 1$, $j=1,2$, since we saw that force noise had a greater impact on numerical stability than sensing noise in our simulations for aLIGO's noise. In Fig. \ref{fig:contour_plots}, we plot the boundary between noise regimes where the numerics converge and the computed covariance matrices are physical (in gray) and those regimes where the numerics fail (either at converging or at generating physical covariance matrices or both) with the available computing resources (in yellow). As a matter of fact, in all the operation regimes in the gray region where the numerics work, the system is entangled. This means that our model predicts optomechanical entanglement in aLIGO, if its classical noise is in this gray region. Recall that the current status of aLIGO corresponds to $\alpha_{F_j} = \alpha_{X_j} = 1$, $j=1,2$, far to the bottom-right in the undecidable yellow region.

If we follow the red dashed line in Fig. \ref{fig:contour_plots}, we continuously sample the force noise spectrum over the boundary where our numerics converge, and the system is entangled, for the partition where we do not trace over the cavity. This boundary corresponds to a lower limit of the maximum amount of classical noise allowed in the system in order to have entanglement. We plot some of these noise curves in Fig. \ref{fig:noise_curves_with_entanglement}. Note that if we set $\alpha_{F_1}$ or $\alpha_{F_2}$ to be unity, the corresponding pairs of $\alpha_{F_1}$, $\alpha_{F_2}$ situated on this boundary would be $\alpha_{F_1} = 1$ and $\alpha_{F_2} \geq 200$, or for $\alpha_{F_2} = 1$ and $\alpha_{F_1} \leq 10^{-10}$. The corresponding noise curves are also plotted in Fig. \ref{fig:noise_curves_with_entanglement}.


\begin{thebibliography}{63}%
\makeatletter
\providecommand \@ifxundefined [1]{%
 \@ifx{#1\undefined}
}%
\providecommand \@ifnum [1]{%
 \ifnum #1\expandafter \@firstoftwo
 \else \expandafter \@secondoftwo
 \fi
}%
\providecommand \@ifx [1]{%
 \ifx #1\expandafter \@firstoftwo
 \else \expandafter \@secondoftwo
 \fi
}%
\providecommand \natexlab [1]{#1}%
\providecommand \enquote  [1]{``#1''}%
\providecommand \bibnamefont  [1]{#1}%
\providecommand \bibfnamefont [1]{#1}%
\providecommand \citenamefont [1]{#1}%
\providecommand \href@noop [0]{\@secondoftwo}%
\providecommand \href [0]{\begingroup \@sanitize@url \@href}%
\providecommand \@href[1]{\@@startlink{#1}\@@href}%
\providecommand \@@href[1]{\endgroup#1\@@endlink}%
\providecommand \@sanitize@url [0]{\catcode `\\12\catcode `\$12\catcode
  `\&12\catcode `\#12\catcode `\^12\catcode `\_12\catcode `\%12\relax}%
\providecommand \@@startlink[1]{}%
\providecommand \@@endlink[0]{}%
\providecommand \url  [0]{\begingroup\@sanitize@url \@url }%
\providecommand \@url [1]{\endgroup\@href {#1}{\urlprefix }}%
\providecommand \urlprefix  [0]{URL }%
\providecommand \Eprint [0]{\href }%
\providecommand \doibase [0]{https://doi.org/}%
\providecommand \selectlanguage [0]{\@gobble}%
\providecommand \bibinfo  [0]{\@secondoftwo}%
\providecommand \bibfield  [0]{\@secondoftwo}%
\providecommand \translation [1]{[#1]}%
\providecommand \BibitemOpen [0]{}%
\providecommand \bibitemStop [0]{}%
\providecommand \bibitemNoStop [0]{.\EOS\space}%
\providecommand \EOS [0]{\spacefactor3000\relax}%
\providecommand \BibitemShut  [1]{\csname bibitem#1\endcsname}%
\let\auto@bib@innerbib\@empty
\bibitem [{\citenamefont {Genes}\ \emph {et~al.}(2009)\citenamefont {Genes},
  \citenamefont {Mari}, \citenamefont {Vitali},\ and\ \citenamefont
  {Tombesi}}]{GENES200933}%
  \BibitemOpen
  \bibfield  {author} {\bibinfo {author} {\bibfnamefont {C.}~\bibnamefont
  {Genes}}, \bibinfo {author} {\bibfnamefont {A.}~\bibnamefont {Mari}},
  \bibinfo {author} {\bibfnamefont {D.}~\bibnamefont {Vitali}},\ and\ \bibinfo
  {author} {\bibfnamefont {P.}~\bibnamefont {Tombesi}},\ }\bibfield  {title}
  {\bibinfo {title} {Chapter 2 quantum effects in optomechanical systems},\
  }in\ \href {https://doi.org/https://doi.org/10.1016/S1049-250X(09)57002-4}
  {\emph {\bibinfo {booktitle} {Advances in Atomic Molecular and Optical
  Physics}}},\ \bibinfo {series} {Advances In Atomic, Molecular, and Optical
  Physics}, Vol.~\bibinfo {volume} {57}\ (\bibinfo  {publisher} {Academic
  Press},\ \bibinfo {address} {San Diego},\ \bibinfo {year} {2009})\ pp.\
  \bibinfo {pages} {33--86}\BibitemShut {NoStop}%
\bibitem [{\citenamefont {Hofer}\ and\ \citenamefont
  {Hammerer}(2015)}]{quantum_control}%
  \BibitemOpen
  \bibfield  {author} {\bibinfo {author} {\bibfnamefont {S.~G.}\ \bibnamefont
  {Hofer}}\ and\ \bibinfo {author} {\bibfnamefont {K.}~\bibnamefont
  {Hammerer}},\ }\bibfield  {title} {\bibinfo {title} {Entanglement-enhanced
  time-continuous quantum control in optomechanics},\ }\href
  {https://doi.org/10.1103/PhysRevA.91.033822} {\bibfield  {journal} {\bibinfo
  {journal} {Phys. Rev. A}\ }\textbf {\bibinfo {volume} {91}},\ \bibinfo
  {pages} {033822} (\bibinfo {year} {2015})}\BibitemShut {NoStop}%
\bibitem [{\citenamefont {Genes}\ \emph {et~al.}(2008)\citenamefont {Genes},
  \citenamefont {Mari}, \citenamefont {Tombesi},\ and\ \citenamefont
  {Vitali}}]{PhysRevA.78.032316}%
  \BibitemOpen
  \bibfield  {author} {\bibinfo {author} {\bibfnamefont {C.}~\bibnamefont
  {Genes}}, \bibinfo {author} {\bibfnamefont {A.}~\bibnamefont {Mari}},
  \bibinfo {author} {\bibfnamefont {P.}~\bibnamefont {Tombesi}},\ and\ \bibinfo
  {author} {\bibfnamefont {D.}~\bibnamefont {Vitali}},\ }\bibfield  {title}
  {\bibinfo {title} {Robust entanglement of a micromechanical resonator with
  output optical fields},\ }\href {https://doi.org/10.1103/PhysRevA.78.032316}
  {\bibfield  {journal} {\bibinfo  {journal} {Phys. Rev. A}\ }\textbf {\bibinfo
  {volume} {78}},\ \bibinfo {pages} {032316} (\bibinfo {year}
  {2008})}\BibitemShut {NoStop}%
\bibitem [{\citenamefont {Miao}\ \emph
  {et~al.}(2010{\natexlab{a}})\citenamefont {Miao}, \citenamefont {Danilishin},
  \citenamefont {M\"{u}ller-Ebhardt},\ and\ \citenamefont {Chen}}]{Miao2010}%
  \BibitemOpen
  \bibfield  {author} {\bibinfo {author} {\bibfnamefont {H.}~\bibnamefont
  {Miao}}, \bibinfo {author} {\bibfnamefont {S.}~\bibnamefont {Danilishin}},
  \bibinfo {author} {\bibfnamefont {H.}~\bibnamefont {M\"{u}ller-Ebhardt}},\
  and\ \bibinfo {author} {\bibfnamefont {Y.}~\bibnamefont {Chen}},\ }\bibfield
  {title} {\bibinfo {title} {Achieving ground state and enhancing
  optomechanical entanglement by recovering information},\ }\href
  {https://doi.org/10.1088/1367-2630/12/8/083032} {\bibfield  {journal}
  {\bibinfo  {journal} {New Journal of Physics}\ }\textbf {\bibinfo {volume}
  {12}},\ \bibinfo {pages} {083032} (\bibinfo {year}
  {2010}{\natexlab{a}})}\BibitemShut {NoStop}%
\bibitem [{\citenamefont {Miao}\ \emph
  {et~al.}(2010{\natexlab{b}})\citenamefont {Miao}, \citenamefont {Danilishin},
  \citenamefont {M\"uller-Ebhardt}, \citenamefont {Rehbein}, \citenamefont
  {Somiya},\ and\ \citenamefont {Chen}}]{PhysRevA.81.012114}%
  \BibitemOpen
  \bibfield  {author} {\bibinfo {author} {\bibfnamefont {H.}~\bibnamefont
  {Miao}}, \bibinfo {author} {\bibfnamefont {S.}~\bibnamefont {Danilishin}},
  \bibinfo {author} {\bibfnamefont {H.}~\bibnamefont {M\"uller-Ebhardt}},
  \bibinfo {author} {\bibfnamefont {H.}~\bibnamefont {Rehbein}}, \bibinfo
  {author} {\bibfnamefont {K.}~\bibnamefont {Somiya}},\ and\ \bibinfo {author}
  {\bibfnamefont {Y.}~\bibnamefont {Chen}},\ }\bibfield  {title} {\bibinfo
  {title} {Probing macroscopic quantum states with a sub-heisenberg accuracy},\
  }\href {https://doi.org/10.1103/PhysRevA.81.012114} {\bibfield  {journal}
  {\bibinfo  {journal} {Phys. Rev. A}\ }\textbf {\bibinfo {volume} {81}},\
  \bibinfo {pages} {012114} (\bibinfo {year} {2010}{\natexlab{b}})}\BibitemShut
  {NoStop}%
\bibitem [{\citenamefont {Miao}\ \emph
  {et~al.}(2010{\natexlab{c}})\citenamefont {Miao}, \citenamefont
  {Danilishin},\ and\ \citenamefont {Chen}}]{universal_quantum_entanglement}%
  \BibitemOpen
  \bibfield  {author} {\bibinfo {author} {\bibfnamefont {H.}~\bibnamefont
  {Miao}}, \bibinfo {author} {\bibfnamefont {S.}~\bibnamefont {Danilishin}},\
  and\ \bibinfo {author} {\bibfnamefont {Y.}~\bibnamefont {Chen}},\ }\bibfield
  {title} {\bibinfo {title} {Universal quantum entanglement between an
  oscillator and continuous fields},\ }\bibfield  {journal} {\bibinfo
  {journal} {Physical Review A}\ }\textbf {\bibinfo {volume} {81}},\ \href
  {https://doi.org/10.1103/physreva.81.052307} {10.1103/physreva.81.052307}
  (\bibinfo {year} {2010}{\natexlab{c}})\BibitemShut {NoStop}%
\bibitem [{\citenamefont {Danilishin}\ \emph {et~al.}(2019)\citenamefont
  {Danilishin}, \citenamefont {Khalili},\ and\ \citenamefont
  {Miao}}]{Danilishin2019}%
  \BibitemOpen
  \bibfield  {author} {\bibinfo {author} {\bibfnamefont {S.~L.}\ \bibnamefont
  {Danilishin}}, \bibinfo {author} {\bibfnamefont {F.~Y.}\ \bibnamefont
  {Khalili}},\ and\ \bibinfo {author} {\bibfnamefont {H.}~\bibnamefont
  {Miao}},\ }\bibfield  {title} {\bibinfo {title} {Advanced quantum techniques
  for future gravitational-wave detectors},\ }\bibfield  {journal} {\bibinfo
  {journal} {Living Reviews in Relativity}\ }\textbf {\bibinfo {volume} {22}},\
  \href {https://doi.org/10.1007/s41114-019-0018-y} {10.1007/s41114-019-0018-y}
  (\bibinfo {year} {2019})\BibitemShut {NoStop}%
\bibitem [{\citenamefont {Hofer}\ \emph {et~al.}(2011)\citenamefont {Hofer},
  \citenamefont {Wieczorek}, \citenamefont {Aspelmeyer},\ and\ \citenamefont
  {Hammerer}}]{Quantum_entanglement_teleportation_pulsed}%
  \BibitemOpen
  \bibfield  {author} {\bibinfo {author} {\bibfnamefont {S.~G.}\ \bibnamefont
  {Hofer}}, \bibinfo {author} {\bibfnamefont {W.}~\bibnamefont {Wieczorek}},
  \bibinfo {author} {\bibfnamefont {M.}~\bibnamefont {Aspelmeyer}},\ and\
  \bibinfo {author} {\bibfnamefont {K.}~\bibnamefont {Hammerer}},\ }\bibfield
  {title} {\bibinfo {title} {Quantum entanglement and teleportation in pulsed
  cavity optomechanics},\ }\href {https://doi.org/10.1103/PhysRevA.84.052327}
  {\bibfield  {journal} {\bibinfo  {journal} {Phys. Rev. A}\ }\textbf {\bibinfo
  {volume} {84}},\ \bibinfo {pages} {052327} (\bibinfo {year}
  {2011})}\BibitemShut {NoStop}%
\bibitem [{\citenamefont {Aspelmeyer}\ \emph {et~al.}(2014)\citenamefont
  {Aspelmeyer}, \citenamefont {Kippenberg},\ and\ \citenamefont
  {Marquardt}}]{Aspelmeyer_cavity_optomechanics}%
  \BibitemOpen
  \bibfield  {author} {\bibinfo {author} {\bibfnamefont {M.}~\bibnamefont
  {Aspelmeyer}}, \bibinfo {author} {\bibfnamefont {T.~J.}\ \bibnamefont
  {Kippenberg}},\ and\ \bibinfo {author} {\bibfnamefont {F.}~\bibnamefont
  {Marquardt}},\ }\bibfield  {title} {\bibinfo {title} {Cavity optomechanics},\
  }\href {https://doi.org/10.1103/revmodphys.86.1391} {\bibfield  {journal}
  {\bibinfo  {journal} {Reviews of Modern Physics}\ }\textbf {\bibinfo {volume}
  {86}},\ \bibinfo {pages} {1391} (\bibinfo {year} {2014})}\BibitemShut
  {NoStop}%
\bibitem [{\citenamefont {Barzanjeh}\ \emph {et~al.}(2019)\citenamefont
  {Barzanjeh}, \citenamefont {Redchenko}, \citenamefont {Peruzzo},
  \citenamefont {Wulf}, \citenamefont {Lewis}, \citenamefont {Arnold},\ and\
  \citenamefont {Fink}}]{entangled_radiation_from_micromechanical}%
  \BibitemOpen
  \bibfield  {author} {\bibinfo {author} {\bibfnamefont {S.}~\bibnamefont
  {Barzanjeh}}, \bibinfo {author} {\bibfnamefont {E.~S.}\ \bibnamefont
  {Redchenko}}, \bibinfo {author} {\bibfnamefont {M.}~\bibnamefont {Peruzzo}},
  \bibinfo {author} {\bibfnamefont {M.}~\bibnamefont {Wulf}}, \bibinfo {author}
  {\bibfnamefont {D.~P.}\ \bibnamefont {Lewis}}, \bibinfo {author}
  {\bibfnamefont {G.}~\bibnamefont {Arnold}},\ and\ \bibinfo {author}
  {\bibfnamefont {J.~M.}\ \bibnamefont {Fink}},\ }\bibfield  {title} {\bibinfo
  {title} {Stationary entangled radiation from micromechanical motion},\ }\href
  {https://doi.org/10.1038/s41586-019-1320-2} {\bibfield  {journal} {\bibinfo
  {journal} {Nature}\ }\textbf {\bibinfo {volume} {570}},\ \bibinfo {pages}
  {480} (\bibinfo {year} {2019})}\BibitemShut {NoStop}%
\bibitem [{\citenamefont {Chen}\ \emph {et~al.}(2020)\citenamefont {Chen},
  \citenamefont {Rossi}, \citenamefont {Mason},\ and\ \citenamefont
  {Schliesser}}]{propagating_optical_modes}%
  \BibitemOpen
  \bibfield  {author} {\bibinfo {author} {\bibfnamefont {J.}~\bibnamefont
  {Chen}}, \bibinfo {author} {\bibfnamefont {M.}~\bibnamefont {Rossi}},
  \bibinfo {author} {\bibfnamefont {D.}~\bibnamefont {Mason}},\ and\ \bibinfo
  {author} {\bibfnamefont {A.}~\bibnamefont {Schliesser}},\ }\bibfield  {title}
  {\bibinfo {title} {Entanglement of propagating optical modes via a mechanical
  interface},\ }\href {https://doi.org/10.1038/s41467-020-14768-1} {\bibfield
  {journal} {\bibinfo  {journal} {Nature Communications}\ }\textbf {\bibinfo
  {volume} {11}},\ \bibinfo {pages} {943} (\bibinfo {year} {2020})}\BibitemShut
  {NoStop}%
\bibitem [{\citenamefont {Riedinger}\ \emph {et~al.}(2018)\citenamefont
  {Riedinger}, \citenamefont {Wallucks}, \citenamefont {Marinkovi{\'{c}}},
  \citenamefont {L{\"o}schnauer}, \citenamefont {Aspelmeyer}, \citenamefont
  {Hong},\ and\ \citenamefont
  {Gr{\"o}blacher}}]{Remote_between_micromechanical}%
  \BibitemOpen
  \bibfield  {author} {\bibinfo {author} {\bibfnamefont {R.}~\bibnamefont
  {Riedinger}}, \bibinfo {author} {\bibfnamefont {A.}~\bibnamefont {Wallucks}},
  \bibinfo {author} {\bibfnamefont {I.}~\bibnamefont {Marinkovi{\'{c}}}},
  \bibinfo {author} {\bibfnamefont {C.}~\bibnamefont {L{\"o}schnauer}},
  \bibinfo {author} {\bibfnamefont {M.}~\bibnamefont {Aspelmeyer}}, \bibinfo
  {author} {\bibfnamefont {S.}~\bibnamefont {Hong}},\ and\ \bibinfo {author}
  {\bibfnamefont {S.}~\bibnamefont {Gr{\"o}blacher}},\ }\bibfield  {title}
  {\bibinfo {title} {Remote quantum entanglement between two micromechanical
  oscillators},\ }\href {https://doi.org/10.1038/s41586-018-0036-z} {\bibfield
  {journal} {\bibinfo  {journal} {Nature}\ }\textbf {\bibinfo {volume} {556}},\
  \bibinfo {pages} {473} (\bibinfo {year} {2018})}\BibitemShut {NoStop}%
\bibitem [{\citenamefont {Ockeloen-Korppi}\ \emph {et~al.}(2018)\citenamefont
  {Ockeloen-Korppi}, \citenamefont {Damsk{\"a}gg}, \citenamefont
  {Pirkkalainen}, \citenamefont {Asjad}, \citenamefont {Clerk}, \citenamefont
  {Massel}, \citenamefont {Woolley},\ and\ \citenamefont
  {Sillanp{\"a}{\"a}}}]{stabilized_entanglement_massive}%
  \BibitemOpen
  \bibfield  {author} {\bibinfo {author} {\bibfnamefont {C.~F.}\ \bibnamefont
  {Ockeloen-Korppi}}, \bibinfo {author} {\bibfnamefont {E.}~\bibnamefont
  {Damsk{\"a}gg}}, \bibinfo {author} {\bibfnamefont {J.-M.}\ \bibnamefont
  {Pirkkalainen}}, \bibinfo {author} {\bibfnamefont {M.}~\bibnamefont {Asjad}},
  \bibinfo {author} {\bibfnamefont {A.~A.}\ \bibnamefont {Clerk}}, \bibinfo
  {author} {\bibfnamefont {F.}~\bibnamefont {Massel}}, \bibinfo {author}
  {\bibfnamefont {M.~J.}\ \bibnamefont {Woolley}},\ and\ \bibinfo {author}
  {\bibfnamefont {M.~A.}\ \bibnamefont {Sillanp{\"a}{\"a}}},\ }\bibfield
  {title} {\bibinfo {title} {Stabilized entanglement of massive mechanical
  oscillators},\ }\href {https://doi.org/10.1038/s41586-018-0038-x} {\bibfield
  {journal} {\bibinfo  {journal} {Nature}\ }\textbf {\bibinfo {volume} {556}},\
  \bibinfo {pages} {478} (\bibinfo {year} {2018})}\BibitemShut {NoStop}%
\bibitem [{\citenamefont {Kotler}\ \emph {et~al.}(2021)\citenamefont {Kotler},
  \citenamefont {Peterson}, \citenamefont {Shojaee}, \citenamefont {Lecocq},
  \citenamefont {Cicak}, \citenamefont {Kwiatkowski}, \citenamefont {Geller},
  \citenamefont {Glancy}, \citenamefont {Knill}, \citenamefont {Simmonds},
  \citenamefont {Aumentado},\ and\ \citenamefont
  {Teufel}}]{Direct_observation_of_macroscopic_entanglement}%
  \BibitemOpen
  \bibfield  {author} {\bibinfo {author} {\bibfnamefont {S.}~\bibnamefont
  {Kotler}}, \bibinfo {author} {\bibfnamefont {G.~A.}\ \bibnamefont
  {Peterson}}, \bibinfo {author} {\bibfnamefont {E.}~\bibnamefont {Shojaee}},
  \bibinfo {author} {\bibfnamefont {F.}~\bibnamefont {Lecocq}}, \bibinfo
  {author} {\bibfnamefont {K.}~\bibnamefont {Cicak}}, \bibinfo {author}
  {\bibfnamefont {A.}~\bibnamefont {Kwiatkowski}}, \bibinfo {author}
  {\bibfnamefont {S.}~\bibnamefont {Geller}}, \bibinfo {author} {\bibfnamefont
  {S.}~\bibnamefont {Glancy}}, \bibinfo {author} {\bibfnamefont
  {E.}~\bibnamefont {Knill}}, \bibinfo {author} {\bibfnamefont {R.~W.}\
  \bibnamefont {Simmonds}}, \bibinfo {author} {\bibfnamefont {J.}~\bibnamefont
  {Aumentado}},\ and\ \bibinfo {author} {\bibfnamefont {J.~D.}\ \bibnamefont
  {Teufel}},\ }\bibfield  {title} {\bibinfo {title} {Direct observation of
  deterministic macroscopic entanglement},\ }\href
  {https://doi.org/10.1126/science.abf2998} {\bibfield  {journal} {\bibinfo
  {journal} {Science}\ }\textbf {\bibinfo {volume} {372}},\ \bibinfo {pages}
  {622} (\bibinfo {year} {2021})}\BibitemShut {NoStop}%
\bibitem [{\citenamefont {Thomas}\ \emph {et~al.}(2021)\citenamefont {Thomas},
  \citenamefont {Parniak}, \citenamefont {{\O}stfeldt}, \citenamefont
  {M{\o}ller}, \citenamefont {B{\ae}rentsen}, \citenamefont {Tsaturyan},
  \citenamefont {Schliesser}, \citenamefont {Appel}, \citenamefont {Zeuthen},\
  and\ \citenamefont {Polzik}}]{macroscopic_mechanical_and_spin}%
  \BibitemOpen
  \bibfield  {author} {\bibinfo {author} {\bibfnamefont {R.~A.}\ \bibnamefont
  {Thomas}}, \bibinfo {author} {\bibfnamefont {M.}~\bibnamefont {Parniak}},
  \bibinfo {author} {\bibfnamefont {C.}~\bibnamefont {{\O}stfeldt}}, \bibinfo
  {author} {\bibfnamefont {C.~B.}\ \bibnamefont {M{\o}ller}}, \bibinfo {author}
  {\bibfnamefont {C.}~\bibnamefont {B{\ae}rentsen}}, \bibinfo {author}
  {\bibfnamefont {Y.}~\bibnamefont {Tsaturyan}}, \bibinfo {author}
  {\bibfnamefont {A.}~\bibnamefont {Schliesser}}, \bibinfo {author}
  {\bibfnamefont {J.}~\bibnamefont {Appel}}, \bibinfo {author} {\bibfnamefont
  {E.}~\bibnamefont {Zeuthen}},\ and\ \bibinfo {author} {\bibfnamefont {E.~S.}\
  \bibnamefont {Polzik}},\ }\bibfield  {title} {\bibinfo {title} {Entanglement
  between distant macroscopic mechanical and spin systems},\ }\href
  {https://doi.org/10.1038/s41567-020-1031-5} {\bibfield  {journal} {\bibinfo
  {journal} {Nature Physics}\ }\textbf {\bibinfo {volume} {17}},\ \bibinfo
  {pages} {228} (\bibinfo {year} {2021})}\BibitemShut {NoStop}%
\bibitem [{\citenamefont {Palomaki}\ \emph {et~al.}(2013)\citenamefont
  {Palomaki}, \citenamefont {Teufel}, \citenamefont {Simmonds},\ and\
  \citenamefont {Lehnert}}]{Mechanical_Motion_with_Microwave_Fields}%
  \BibitemOpen
  \bibfield  {author} {\bibinfo {author} {\bibfnamefont {T.~A.}\ \bibnamefont
  {Palomaki}}, \bibinfo {author} {\bibfnamefont {J.~D.}\ \bibnamefont
  {Teufel}}, \bibinfo {author} {\bibfnamefont {R.~W.}\ \bibnamefont
  {Simmonds}},\ and\ \bibinfo {author} {\bibfnamefont {K.~W.}\ \bibnamefont
  {Lehnert}},\ }\bibfield  {title} {\bibinfo {title} {Entangling mechanical
  motion with microwave fields},\ }\href
  {https://doi.org/10.1126/science.1244563} {\bibfield  {journal} {\bibinfo
  {journal} {Science}\ }\textbf {\bibinfo {volume} {342}},\ \bibinfo {pages}
  {710} (\bibinfo {year} {2013})},\ \Eprint
  {https://arxiv.org/abs/https://www.science.org/doi/pdf/10.1126/science.1244563}
  {https://www.science.org/doi/pdf/10.1126/science.1244563} \BibitemShut
  {NoStop}%
\bibitem [{\citenamefont {Romero-Isart}(2011)}]{PhysRevA.84.052121}%
  \BibitemOpen
  \bibfield  {author} {\bibinfo {author} {\bibfnamefont {O.}~\bibnamefont
  {Romero-Isart}},\ }\bibfield  {title} {\bibinfo {title} {Quantum
  superposition of massive objects and collapse models},\ }\href
  {https://doi.org/10.1103/PhysRevA.84.052121} {\bibfield  {journal} {\bibinfo
  {journal} {Phys. Rev. A}\ }\textbf {\bibinfo {volume} {84}},\ \bibinfo
  {pages} {052121} (\bibinfo {year} {2011})}\BibitemShut {NoStop}%
\bibitem [{\citenamefont {M\"uller-Ebhardt}\ \emph {et~al.}(2008)\citenamefont
  {M\"uller-Ebhardt}, \citenamefont {Rehbein}, \citenamefont {Schnabel},
  \citenamefont {Danzmann},\ and\ \citenamefont
  {Chen}}]{entanglement_macroscopic}%
  \BibitemOpen
  \bibfield  {author} {\bibinfo {author} {\bibfnamefont {H.}~\bibnamefont
  {M\"uller-Ebhardt}}, \bibinfo {author} {\bibfnamefont {H.}~\bibnamefont
  {Rehbein}}, \bibinfo {author} {\bibfnamefont {R.}~\bibnamefont {Schnabel}},
  \bibinfo {author} {\bibfnamefont {K.}~\bibnamefont {Danzmann}},\ and\
  \bibinfo {author} {\bibfnamefont {Y.}~\bibnamefont {Chen}},\ }\bibfield
  {title} {\bibinfo {title} {Entanglement of macroscopic test masses and the
  standard quantum limit in laser interferometry},\ }\href
  {https://doi.org/10.1103/PhysRevLett.100.013601} {\bibfield  {journal}
  {\bibinfo  {journal} {Phys. Rev. Lett.}\ }\textbf {\bibinfo {volume} {100}},\
  \bibinfo {pages} {013601} (\bibinfo {year} {2008})}\BibitemShut {NoStop}%
\bibitem [{\citenamefont {Schnabel}(2015)}]{EPR_entangled_motion_of}%
  \BibitemOpen
  \bibfield  {author} {\bibinfo {author} {\bibfnamefont {R.}~\bibnamefont
  {Schnabel}},\ }\bibfield  {title} {\bibinfo {title}
  {Einstein-podolsky-rosen--entangled motion of two massive objects},\ }\href
  {https://doi.org/10.1103/PhysRevA.92.012126} {\bibfield  {journal} {\bibinfo
  {journal} {Phys. Rev. A}\ }\textbf {\bibinfo {volume} {92}},\ \bibinfo
  {pages} {012126} (\bibinfo {year} {2015})}\BibitemShut {NoStop}%
\bibitem [{\citenamefont {Wang}\ \emph {et~al.}(2016)\citenamefont {Wang},
  \citenamefont {L\"u}, \citenamefont {Wang}, \citenamefont {You},\ and\
  \citenamefont {Wu}}]{Macroscopic_quantum_entanglement_modulated}%
  \BibitemOpen
  \bibfield  {author} {\bibinfo {author} {\bibfnamefont {M.}~\bibnamefont
  {Wang}}, \bibinfo {author} {\bibfnamefont {X.-Y.}\ \bibnamefont {L\"u}},
  \bibinfo {author} {\bibfnamefont {Y.-D.}\ \bibnamefont {Wang}}, \bibinfo
  {author} {\bibfnamefont {J.~Q.}\ \bibnamefont {You}},\ and\ \bibinfo {author}
  {\bibfnamefont {Y.}~\bibnamefont {Wu}},\ }\bibfield  {title} {\bibinfo
  {title} {Macroscopic quantum entanglement in modulated optomechanics},\
  }\href {https://doi.org/10.1103/PhysRevA.94.053807} {\bibfield  {journal}
  {\bibinfo  {journal} {Phys. Rev. A}\ }\textbf {\bibinfo {volume} {94}},\
  \bibinfo {pages} {053807} (\bibinfo {year} {2016})}\BibitemShut {NoStop}%
\bibitem [{\citenamefont {Pirandola}\ \emph {et~al.}(2006)\citenamefont
  {Pirandola}, \citenamefont {Vitali}, \citenamefont {Tombesi},\ and\
  \citenamefont {Lloyd}}]{Pirandola_Entanglement_Swapping}%
  \BibitemOpen
  \bibfield  {author} {\bibinfo {author} {\bibfnamefont {S.}~\bibnamefont
  {Pirandola}}, \bibinfo {author} {\bibfnamefont {D.}~\bibnamefont {Vitali}},
  \bibinfo {author} {\bibfnamefont {P.}~\bibnamefont {Tombesi}},\ and\ \bibinfo
  {author} {\bibfnamefont {S.}~\bibnamefont {Lloyd}},\ }\bibfield  {title}
  {\bibinfo {title} {Macroscopic entanglement by entanglement swapping},\
  }\href {https://doi.org/10.1103/PhysRevLett.97.150403} {\bibfield  {journal}
  {\bibinfo  {journal} {Phys. Rev. Lett.}\ }\textbf {\bibinfo {volume} {97}},\
  \bibinfo {pages} {150403} (\bibinfo {year} {2006})}\BibitemShut {NoStop}%
\bibitem [{\citenamefont {Ludwig}\ \emph {et~al.}(2010)\citenamefont {Ludwig},
  \citenamefont {Hammerer},\ and\ \citenamefont {Marquardt}}]{ludwig_2010}%
  \BibitemOpen
  \bibfield  {author} {\bibinfo {author} {\bibfnamefont {M.}~\bibnamefont
  {Ludwig}}, \bibinfo {author} {\bibfnamefont {K.}~\bibnamefont {Hammerer}},\
  and\ \bibinfo {author} {\bibfnamefont {F.}~\bibnamefont {Marquardt}},\
  }\bibfield  {title} {\bibinfo {title} {Entanglement of mechanical oscillators
  coupled to a nonequilibrium environment},\ }\href
  {https://doi.org/10.1103/PhysRevA.82.012333} {\bibfield  {journal} {\bibinfo
  {journal} {Phys. Rev. A}\ }\textbf {\bibinfo {volume} {82}},\ \bibinfo
  {pages} {012333} (\bibinfo {year} {2010})}\BibitemShut {NoStop}%
\bibitem [{\citenamefont {Gut}\ \emph {et~al.}(2020)\citenamefont {Gut},
  \citenamefont {Winkler}, \citenamefont {Hoelscher-Obermaier}, \citenamefont
  {Hofer}, \citenamefont {Nia}, \citenamefont {Walk}, \citenamefont {Steffens},
  \citenamefont {Eisert}, \citenamefont {Wieczorek}, \citenamefont {Slater},
  \citenamefont {Aspelmeyer},\ and\ \citenamefont
  {Hammerer}}]{corentin_klemens_stat_optomech_entanglement}%
  \BibitemOpen
  \bibfield  {author} {\bibinfo {author} {\bibfnamefont {C.}~\bibnamefont
  {Gut}}, \bibinfo {author} {\bibfnamefont {K.}~\bibnamefont {Winkler}},
  \bibinfo {author} {\bibfnamefont {J.}~\bibnamefont {Hoelscher-Obermaier}},
  \bibinfo {author} {\bibfnamefont {S.~G.}\ \bibnamefont {Hofer}}, \bibinfo
  {author} {\bibfnamefont {R.~M.}\ \bibnamefont {Nia}}, \bibinfo {author}
  {\bibfnamefont {N.}~\bibnamefont {Walk}}, \bibinfo {author} {\bibfnamefont
  {A.}~\bibnamefont {Steffens}}, \bibinfo {author} {\bibfnamefont
  {J.}~\bibnamefont {Eisert}}, \bibinfo {author} {\bibfnamefont
  {W.}~\bibnamefont {Wieczorek}}, \bibinfo {author} {\bibfnamefont {J.~A.}\
  \bibnamefont {Slater}}, \bibinfo {author} {\bibfnamefont {M.}~\bibnamefont
  {Aspelmeyer}},\ and\ \bibinfo {author} {\bibfnamefont {K.}~\bibnamefont
  {Hammerer}},\ }\bibfield  {title} {\bibinfo {title} {Stationary
  optomechanical entanglement between a mechanical oscillator and its
  measurement apparatus},\ }\href
  {https://doi.org/10.1103/PhysRevResearch.2.033244} {\bibfield  {journal}
  {\bibinfo  {journal} {Phys. Rev. Res.}\ }\textbf {\bibinfo {volume} {2}},\
  \bibinfo {pages} {033244} (\bibinfo {year} {2020})}\BibitemShut {NoStop}%
\bibitem [{\citenamefont {Vitali}\ \emph {et~al.}(2007)\citenamefont {Vitali},
  \citenamefont {Gigan}, \citenamefont {Ferreira}, \citenamefont {B\"ohm},
  \citenamefont {Tombesi}, \citenamefont {Guerreiro}, \citenamefont {Vedral},
  \citenamefont {Zeilinger},\ and\ \citenamefont
  {Aspelmeyer}}]{PhysRevLett.98.030405}%
  \BibitemOpen
  \bibfield  {author} {\bibinfo {author} {\bibfnamefont {D.}~\bibnamefont
  {Vitali}}, \bibinfo {author} {\bibfnamefont {S.}~\bibnamefont {Gigan}},
  \bibinfo {author} {\bibfnamefont {A.}~\bibnamefont {Ferreira}}, \bibinfo
  {author} {\bibfnamefont {H.~R.}\ \bibnamefont {B\"ohm}}, \bibinfo {author}
  {\bibfnamefont {P.}~\bibnamefont {Tombesi}}, \bibinfo {author} {\bibfnamefont
  {A.}~\bibnamefont {Guerreiro}}, \bibinfo {author} {\bibfnamefont
  {V.}~\bibnamefont {Vedral}}, \bibinfo {author} {\bibfnamefont
  {A.}~\bibnamefont {Zeilinger}},\ and\ \bibinfo {author} {\bibfnamefont
  {M.}~\bibnamefont {Aspelmeyer}},\ }\bibfield  {title} {\bibinfo {title}
  {Optomechanical entanglement between a movable mirror and a cavity field},\
  }\href {https://doi.org/10.1103/PhysRevLett.98.030405} {\bibfield  {journal}
  {\bibinfo  {journal} {Phys. Rev. Lett.}\ }\textbf {\bibinfo {volume} {98}},\
  \bibinfo {pages} {030405} (\bibinfo {year} {2007})}\BibitemShut {NoStop}%
\bibitem [{\citenamefont {Paternostro}\ \emph {et~al.}(2007)\citenamefont
  {Paternostro}, \citenamefont {Vitali}, \citenamefont {Gigan}, \citenamefont
  {Kim}, \citenamefont {Brukner}, \citenamefont {Eisert},\ and\ \citenamefont
  {Aspelmeyer}}]{PhysRevLett.99.250401}%
  \BibitemOpen
  \bibfield  {author} {\bibinfo {author} {\bibfnamefont {M.}~\bibnamefont
  {Paternostro}}, \bibinfo {author} {\bibfnamefont {D.}~\bibnamefont {Vitali}},
  \bibinfo {author} {\bibfnamefont {S.}~\bibnamefont {Gigan}}, \bibinfo
  {author} {\bibfnamefont {M.~S.}\ \bibnamefont {Kim}}, \bibinfo {author}
  {\bibfnamefont {C.}~\bibnamefont {Brukner}}, \bibinfo {author} {\bibfnamefont
  {J.}~\bibnamefont {Eisert}},\ and\ \bibinfo {author} {\bibfnamefont
  {M.}~\bibnamefont {Aspelmeyer}},\ }\bibfield  {title} {\bibinfo {title}
  {Creating and probing multipartite macroscopic entanglement with light},\
  }\href {https://doi.org/10.1103/PhysRevLett.99.250401} {\bibfield  {journal}
  {\bibinfo  {journal} {Phys. Rev. Lett.}\ }\textbf {\bibinfo {volume} {99}},\
  \bibinfo {pages} {250401} (\bibinfo {year} {2007})}\BibitemShut {NoStop}%
\bibitem [{\citenamefont {Giovannetti}\ and\ \citenamefont
  {Vitali}(2001)}]{PhysRevA.63.023812}%
  \BibitemOpen
  \bibfield  {author} {\bibinfo {author} {\bibfnamefont {V.}~\bibnamefont
  {Giovannetti}}\ and\ \bibinfo {author} {\bibfnamefont {D.}~\bibnamefont
  {Vitali}},\ }\bibfield  {title} {\bibinfo {title} {Phase-noise measurement in
  a cavity with a movable mirror undergoing quantum brownian motion},\ }\href
  {https://doi.org/10.1103/PhysRevA.63.023812} {\bibfield  {journal} {\bibinfo
  {journal} {Phys. Rev. A}\ }\textbf {\bibinfo {volume} {63}},\ \bibinfo
  {pages} {023812} (\bibinfo {year} {2001})}\BibitemShut {NoStop}%
\bibitem [{\citenamefont {Saulson}(1990)}]{structural_damping}%
  \BibitemOpen
  \bibfield  {author} {\bibinfo {author} {\bibfnamefont {P.~R.}\ \bibnamefont
  {Saulson}},\ }\bibfield  {title} {\bibinfo {title} {Thermal noise in
  mechanical experiments},\ }\href {https://doi.org/10.1103/PhysRevD.42.2437}
  {\bibfield  {journal} {\bibinfo  {journal} {Phys. Rev. D}\ }\textbf {\bibinfo
  {volume} {42}},\ \bibinfo {pages} {2437} (\bibinfo {year}
  {1990})}\BibitemShut {NoStop}%
\bibitem [{\citenamefont {Neben}\ \emph {et~al.}(2012)\citenamefont {Neben},
  \citenamefont {Bodiya}, \citenamefont {Wipf}, \citenamefont {Oelker},
  \citenamefont {Corbitt},\ and\ \citenamefont
  {Mavalvala}}]{structural_thermal_mirror}%
  \BibitemOpen
  \bibfield  {author} {\bibinfo {author} {\bibfnamefont {A.~R.}\ \bibnamefont
  {Neben}}, \bibinfo {author} {\bibfnamefont {T.~P.}\ \bibnamefont {Bodiya}},
  \bibinfo {author} {\bibfnamefont {C.}~\bibnamefont {Wipf}}, \bibinfo {author}
  {\bibfnamefont {E.}~\bibnamefont {Oelker}}, \bibinfo {author} {\bibfnamefont
  {T.}~\bibnamefont {Corbitt}},\ and\ \bibinfo {author} {\bibfnamefont
  {N.}~\bibnamefont {Mavalvala}},\ }\bibfield  {title} {\bibinfo {title}
  {Structural thermal noise in gram-scale mirror oscillators},\ }\href
  {https://doi.org/10.1088/1367-2630/14/11/115008} {\bibfield  {journal}
  {\bibinfo  {journal} {New Journal of Physics}\ }\textbf {\bibinfo {volume}
  {14}},\ \bibinfo {pages} {115008} (\bibinfo {year} {2012})}\BibitemShut
  {NoStop}%
\bibitem [{\citenamefont {Fedorov}\ \emph {et~al.}(2018)\citenamefont
  {Fedorov}, \citenamefont {Sudhir}, \citenamefont {Schilling}, \citenamefont
  {Schütz}, \citenamefont {Wilson},\ and\ \citenamefont
  {Kippenberg}}]{Evidence_for_structural_damping}%
  \BibitemOpen
  \bibfield  {author} {\bibinfo {author} {\bibfnamefont {S.}~\bibnamefont
  {Fedorov}}, \bibinfo {author} {\bibfnamefont {V.}~\bibnamefont {Sudhir}},
  \bibinfo {author} {\bibfnamefont {R.}~\bibnamefont {Schilling}}, \bibinfo
  {author} {\bibfnamefont {H.}~\bibnamefont {Schütz}}, \bibinfo {author}
  {\bibfnamefont {D.}~\bibnamefont {Wilson}},\ and\ \bibinfo {author}
  {\bibfnamefont {T.}~\bibnamefont {Kippenberg}},\ }\bibfield  {title}
  {\bibinfo {title} {Evidence for structural damping in a high-stress silicon
  nitride nanobeam and its implications for quantum optomechanics},\ }\href
  {https://doi.org/https://doi.org/10.1016/j.physleta.2017.05.046} {\bibfield
  {journal} {\bibinfo  {journal} {Physics Letters A}\ }\textbf {\bibinfo
  {volume} {382}},\ \bibinfo {pages} {2251} (\bibinfo {year} {2018})},\
  \bibinfo {note} {special Issue in memory of Professor V.B.
  Braginsky}\BibitemShut {NoStop}%
\bibitem [{\citenamefont {Gr{\"o}blacher}\ \emph {et~al.}(2015)\citenamefont
  {Gr{\"o}blacher}, \citenamefont {Trubarov}, \citenamefont {Prigge},
  \citenamefont {Cole}, \citenamefont {Aspelmeyer},\ and\ \citenamefont
  {Eisert}}]{non_markovian_micromech_brownian}%
  \BibitemOpen
  \bibfield  {author} {\bibinfo {author} {\bibfnamefont {S.}~\bibnamefont
  {Gr{\"o}blacher}}, \bibinfo {author} {\bibfnamefont {A.}~\bibnamefont
  {Trubarov}}, \bibinfo {author} {\bibfnamefont {N.}~\bibnamefont {Prigge}},
  \bibinfo {author} {\bibfnamefont {G.~D.}\ \bibnamefont {Cole}}, \bibinfo
  {author} {\bibfnamefont {M.}~\bibnamefont {Aspelmeyer}},\ and\ \bibinfo
  {author} {\bibfnamefont {J.}~\bibnamefont {Eisert}},\ }\bibfield  {title}
  {\bibinfo {title} {Observation of non-markovian micromechanical brownian
  motion},\ }\href {https://doi.org/10.1038/ncomms8606} {\bibfield  {journal}
  {\bibinfo  {journal} {Nature Communications}\ }\textbf {\bibinfo {volume}
  {6}},\ \bibinfo {pages} {7606} (\bibinfo {year} {2015})}\BibitemShut
  {NoStop}%
\bibitem [{\citenamefont {Aasi}\ \emph {et~al.}(2015)\citenamefont {Aasi} \emph
  {et~al.}}]{adv_ligo}%
  \BibitemOpen
  \bibfield  {author} {\bibinfo {author} {\bibfnamefont {J.}~\bibnamefont
  {Aasi}} \emph {et~al.},\ }\bibfield  {title} {\bibinfo {title} {Advanced
  ligo},\ }\href {https://doi.org/10.1088/0264-9381/32/7/074001} {\bibfield
  {journal} {\bibinfo  {journal} {Classical and Quantum Gravity}\ }\textbf
  {\bibinfo {volume} {32}},\ \bibinfo {pages} {074001} (\bibinfo {year}
  {2015})}\BibitemShut {NoStop}%
\bibitem [{\citenamefont {Yu}\ and\ \citenamefont {members of~the LIGO
  Scientific~Collaboration}(2020)}]{correlations_kilogram_mass_mirrors}%
  \BibitemOpen
  \bibfield  {author} {\bibinfo {author} {\bibfnamefont {H.}~\bibnamefont
  {Yu}}\ and\ \bibinfo {author} {\bibnamefont {members of~the LIGO
  Scientific~Collaboration}},\ }\bibfield  {title} {\bibinfo {title} {Quantum
  correlations between light and the kilogram-mass mirrors of ligo},\ }\href
  {https://doi.org/10.1038/s41586-020-2420-8} {\bibfield  {journal} {\bibinfo
  {journal} {Nature}\ }\textbf {\bibinfo {volume} {583}},\ \bibinfo {pages}
  {43} (\bibinfo {year} {2020})}\BibitemShut {NoStop}%
\bibitem [{Note1()}]{Note1}%
  \BibitemOpen
  \bibinfo {note} {The quantum noise's beating of the SQL is only inferred by
  subtracting a classical noise floor that was obtained through
  calibration}\BibitemShut {NoStop}%
\bibitem [{\citenamefont {Buonanno}\ and\ \citenamefont
  {Chen}(2003)}]{scaling_law}%
  \BibitemOpen
  \bibfield  {author} {\bibinfo {author} {\bibfnamefont {A.}~\bibnamefont
  {Buonanno}}\ and\ \bibinfo {author} {\bibfnamefont {Y.}~\bibnamefont
  {Chen}},\ }\bibfield  {title} {\bibinfo {title} {Scaling law in signal
  recycled laser-interferometer gravitational-wave detectors},\ }\bibfield
  {journal} {\bibinfo  {journal} {Physical Review D}\ }\textbf {\bibinfo
  {volume} {67}},\ \href {https://doi.org/10.1103/physrevd.67.062002}
  {10.1103/physrevd.67.062002} (\bibinfo {year} {2003})\BibitemShut {NoStop}%
\bibitem [{\citenamefont {Chen}(2013)}]{Chen_2013}%
  \BibitemOpen
  \bibfield  {author} {\bibinfo {author} {\bibfnamefont {Y.}~\bibnamefont
  {Chen}},\ }\bibfield  {title} {\bibinfo {title} {Macroscopic quantum
  mechanics: theory and experimental concepts of optomechanics},\ }\href
  {https://doi.org/10.1088/0953-4075/46/10/104001} {\bibfield  {journal}
  {\bibinfo  {journal} {Journal of Physics B: Atomic, Molecular and Optical
  Physics}\ }\textbf {\bibinfo {volume} {46}},\ \bibinfo {pages} {104001}
  (\bibinfo {year} {2013})}\BibitemShut {NoStop}%
\bibitem [{Note2()}]{Note2}%
  \BibitemOpen
  \bibinfo {note} {We use the convention $\protect \mathcal {F}\{f(t)\} =
  \DOTSI \intop \ilimits@ _{-\infty }^{\infty } f(t)e^{i\omega t}
  dt$}\BibitemShut {NoStop}%
\bibitem [{\citenamefont {Caves}\ and\ \citenamefont
  {Schumaker}(1985)}]{caves_two_photon}%
  \BibitemOpen
  \bibfield  {author} {\bibinfo {author} {\bibfnamefont {C.~M.}\ \bibnamefont
  {Caves}}\ and\ \bibinfo {author} {\bibfnamefont {B.~L.}\ \bibnamefont
  {Schumaker}},\ }\bibfield  {title} {\bibinfo {title} {New formalism for
  two-photon quantum optics. i. quadrature phases and squeezed states},\ }\href
  {https://doi.org/10.1103/PhysRevA.31.3068} {\bibfield  {journal} {\bibinfo
  {journal} {Phys. Rev. A}\ }\textbf {\bibinfo {volume} {31}},\ \bibinfo
  {pages} {3068} (\bibinfo {year} {1985})}\BibitemShut {NoStop}%
\bibitem [{\citenamefont {Schumaker}\ and\ \citenamefont
  {Caves}(1985)}]{caves_two_photon_2}%
  \BibitemOpen
  \bibfield  {author} {\bibinfo {author} {\bibfnamefont {B.~L.}\ \bibnamefont
  {Schumaker}}\ and\ \bibinfo {author} {\bibfnamefont {C.~M.}\ \bibnamefont
  {Caves}},\ }\bibfield  {title} {\bibinfo {title} {New formalism for
  two-photon quantum optics. ii. mathematical foundation and compact
  notation},\ }\href {https://doi.org/10.1103/PhysRevA.31.3093} {\bibfield
  {journal} {\bibinfo  {journal} {Phys. Rev. A}\ }\textbf {\bibinfo {volume}
  {31}},\ \bibinfo {pages} {3093} (\bibinfo {year} {1985})}\BibitemShut
  {NoStop}%
\bibitem [{\citenamefont {Danilishin}\ and\ \citenamefont
  {Khalili}(2012)}]{Danilishin_quantum_meas_theory}%
  \BibitemOpen
  \bibfield  {author} {\bibinfo {author} {\bibfnamefont {S.~L.}\ \bibnamefont
  {Danilishin}}\ and\ \bibinfo {author} {\bibfnamefont {F.~Y.}\ \bibnamefont
  {Khalili}},\ }\bibfield  {title} {\bibinfo {title} {Quantum measurement
  theory in gravitational-wave detectors},\ }\bibfield  {journal} {\bibinfo
  {journal} {Living Reviews in Relativity}\ }\textbf {\bibinfo {volume} {15}},\
  \href {https://doi.org/10.12942/lrr-2012-5} {10.12942/lrr-2012-5} (\bibinfo
  {year} {2012})\BibitemShut {NoStop}%
\bibitem [{\citenamefont {Kubo}(1966)}]{kubo1966fluctuation}%
  \BibitemOpen
  \bibfield  {author} {\bibinfo {author} {\bibfnamefont {R.}~\bibnamefont
  {Kubo}},\ }\bibfield  {title} {\bibinfo {title} {The fluctuation-dissipation
  theorem},\ }\href@noop {} {\bibfield  {journal} {\bibinfo  {journal} {Reports
  on progress in physics}\ }\textbf {\bibinfo {volume} {29}},\ \bibinfo {pages}
  {255} (\bibinfo {year} {1966})}\BibitemShut {NoStop}%
\bibitem [{\citenamefont {Kimble}\ \emph {et~al.}(2001)\citenamefont {Kimble},
  \citenamefont {Levin}, \citenamefont {Matsko}, \citenamefont {Thorne},\ and\
  \citenamefont {Vyatchanin}}]{Kimble_conversion_of_conventional}%
  \BibitemOpen
  \bibfield  {author} {\bibinfo {author} {\bibfnamefont {H.~J.}\ \bibnamefont
  {Kimble}}, \bibinfo {author} {\bibfnamefont {Y.}~\bibnamefont {Levin}},
  \bibinfo {author} {\bibfnamefont {A.~B.}\ \bibnamefont {Matsko}}, \bibinfo
  {author} {\bibfnamefont {K.~S.}\ \bibnamefont {Thorne}},\ and\ \bibinfo
  {author} {\bibfnamefont {S.~P.}\ \bibnamefont {Vyatchanin}},\ }\bibfield
  {title} {\bibinfo {title} {Conversion of conventional gravitational-wave
  interferometers into quantum nondemolition interferometers by modifying their
  input and/or output optics},\ }\bibfield  {journal} {\bibinfo  {journal}
  {Physical Review D}\ }\textbf {\bibinfo {volume} {65}},\ \href
  {https://doi.org/10.1103/physrevd.65.022002} {10.1103/physrevd.65.022002}
  (\bibinfo {year} {2001})\BibitemShut {NoStop}%
\bibitem [{\citenamefont {Peres}(1996)}]{peres_ppt}%
  \BibitemOpen
  \bibfield  {author} {\bibinfo {author} {\bibfnamefont {A.}~\bibnamefont
  {Peres}},\ }\bibfield  {title} {\bibinfo {title} {Separability criterion for
  density matrices},\ }\href {https://doi.org/10.1103/PhysRevLett.77.1413}
  {\bibfield  {journal} {\bibinfo  {journal} {Phys. Rev. Lett.}\ }\textbf
  {\bibinfo {volume} {77}},\ \bibinfo {pages} {1413} (\bibinfo {year}
  {1996})}\BibitemShut {NoStop}%
\bibitem [{\citenamefont {Adesso}\ and\ \citenamefont
  {Illuminati}(2007)}]{Adesso_2007}%
  \BibitemOpen
  \bibfield  {author} {\bibinfo {author} {\bibfnamefont {G.}~\bibnamefont
  {Adesso}}\ and\ \bibinfo {author} {\bibfnamefont {F.}~\bibnamefont
  {Illuminati}},\ }\bibfield  {title} {\bibinfo {title} {Entanglement in
  continuous-variable systems: recent advances and current perspectives},\
  }\href {https://doi.org/10.1088/1751-8113/40/28/s01} {\bibfield  {journal}
  {\bibinfo  {journal} {Journal of Physics A: Mathematical and Theoretical}\
  }\textbf {\bibinfo {volume} {40}},\ \bibinfo {pages} {7821} (\bibinfo {year}
  {2007})}\BibitemShut {NoStop}%
\bibitem [{\citenamefont {Werner}\ and\ \citenamefont
  {Wolf}(2001)}]{bound_entangled_states}%
  \BibitemOpen
  \bibfield  {author} {\bibinfo {author} {\bibfnamefont {R.~F.}\ \bibnamefont
  {Werner}}\ and\ \bibinfo {author} {\bibfnamefont {M.~M.}\ \bibnamefont
  {Wolf}},\ }\bibfield  {title} {\bibinfo {title} {Bound entangled gaussian
  states},\ }\href {https://doi.org/10.1103/PhysRevLett.86.3658} {\bibfield
  {journal} {\bibinfo  {journal} {Phys. Rev. Lett.}\ }\textbf {\bibinfo
  {volume} {86}},\ \bibinfo {pages} {3658} (\bibinfo {year}
  {2001})}\BibitemShut {NoStop}%
\bibitem [{\citenamefont {Simon}(2000)}]{simon}%
  \BibitemOpen
  \bibfield  {author} {\bibinfo {author} {\bibfnamefont {R.}~\bibnamefont
  {Simon}},\ }\bibfield  {title} {\bibinfo {title} {Peres-horodecki
  separability criterion for continuous variable systems},\ }\href
  {https://doi.org/10.1103/PhysRevLett.84.2726} {\bibfield  {journal} {\bibinfo
   {journal} {Phys. Rev. Lett.}\ }\textbf {\bibinfo {volume} {84}},\ \bibinfo
  {pages} {2726} (\bibinfo {year} {2000})}\BibitemShut {NoStop}%
\bibitem [{\citenamefont {Vidal}\ and\ \citenamefont
  {Werner}(2002)}]{negativity}%
  \BibitemOpen
  \bibfield  {author} {\bibinfo {author} {\bibfnamefont {G.}~\bibnamefont
  {Vidal}}\ and\ \bibinfo {author} {\bibfnamefont {R.~F.}\ \bibnamefont
  {Werner}},\ }\bibfield  {title} {\bibinfo {title} {Computable measure of
  entanglement},\ }\href {https://doi.org/10.1103/PhysRevA.65.032314}
  {\bibfield  {journal} {\bibinfo  {journal} {Phys. Rev. A}\ }\textbf {\bibinfo
  {volume} {65}},\ \bibinfo {pages} {032314} (\bibinfo {year}
  {2002})}\BibitemShut {NoStop}%
\bibitem [{\citenamefont {Serafini}(2017)}]{Serafini2017}%
  \BibitemOpen
  \bibfield  {author} {\bibinfo {author} {\bibfnamefont {A.}~\bibnamefont
  {Serafini}},\ }\href {https://doi.org/10.1201/9781315118727} {\emph {\bibinfo
  {title} {Quantum Continuous Variables}}}\ (\bibinfo  {publisher} {{CRC}
  Press},\ \bibinfo {year} {2017})\BibitemShut {NoStop}%
\bibitem [{\citenamefont {Somiya}\ \emph {et~al.}(2012)\citenamefont {Somiya}
  \emph {et~al.}}]{KAGRA}%
  \BibitemOpen
  \bibfield  {author} {\bibinfo {author} {\bibfnamefont {K.}~\bibnamefont
  {Somiya}} \emph {et~al.},\ }\bibfield  {title} {\bibinfo {title} {Detector
  configuration of kagra–the japanese cryogenic gravitational-wave
  detector},\ }\href {https://doi.org/10.1088/0264-9381/29/12/124007}
  {\bibfield  {journal} {\bibinfo  {journal} {Classical and Quantum Gravity}\
  }\textbf {\bibinfo {volume} {29}},\ \bibinfo {pages} {124007} (\bibinfo
  {year} {2012})}\BibitemShut {NoStop}%
\bibitem [{\citenamefont {Acernese}\ \emph {et~al.}(2014)\citenamefont
  {Acernese} \emph {et~al.}}]{VIRGO}%
  \BibitemOpen
  \bibfield  {author} {\bibinfo {author} {\bibfnamefont {F.}~\bibnamefont
  {Acernese}} \emph {et~al.},\ }\bibfield  {title} {\bibinfo {title} {Advanced
  virgo: a second-generation interferometric gravitational wave detector},\
  }\href {https://doi.org/10.1088/0264-9381/32/2/024001} {\bibfield  {journal}
  {\bibinfo  {journal} {Classical and Quantum Gravity}\ }\textbf {\bibinfo
  {volume} {32}},\ \bibinfo {pages} {024001} (\bibinfo {year}
  {2014})}\BibitemShut {NoStop}%
\bibitem [{\citenamefont {Cripe}\ \emph {et~al.}(2019)\citenamefont {Cripe},
  \citenamefont {Aggarwal}, \citenamefont {Lanza}, \citenamefont {Libson},
  \citenamefont {Singh}, \citenamefont {Heu}, \citenamefont {Follman},
  \citenamefont {Cole}, \citenamefont {Mavalvala},\ and\ \citenamefont
  {Corbitt}}]{thomas_corbitt}%
  \BibitemOpen
  \bibfield  {author} {\bibinfo {author} {\bibfnamefont {J.}~\bibnamefont
  {Cripe}}, \bibinfo {author} {\bibfnamefont {N.}~\bibnamefont {Aggarwal}},
  \bibinfo {author} {\bibfnamefont {R.}~\bibnamefont {Lanza}}, \bibinfo
  {author} {\bibfnamefont {A.}~\bibnamefont {Libson}}, \bibinfo {author}
  {\bibfnamefont {R.}~\bibnamefont {Singh}}, \bibinfo {author} {\bibfnamefont
  {P.}~\bibnamefont {Heu}}, \bibinfo {author} {\bibfnamefont {D.}~\bibnamefont
  {Follman}}, \bibinfo {author} {\bibfnamefont {G.~D.}\ \bibnamefont {Cole}},
  \bibinfo {author} {\bibfnamefont {N.}~\bibnamefont {Mavalvala}},\ and\
  \bibinfo {author} {\bibfnamefont {T.}~\bibnamefont {Corbitt}},\ }\bibfield
  {title} {\bibinfo {title} {Measurement of quantum back action in the audio
  band at room temperature},\ }\href
  {https://doi.org/10.1038/s41586-019-1051-4} {\bibfield  {journal} {\bibinfo
  {journal} {Nature}\ }\textbf {\bibinfo {volume} {568}},\ \bibinfo {pages}
  {364} (\bibinfo {year} {2019})}\BibitemShut {NoStop}%
\bibitem [{\citenamefont {Komori}\ \emph {et~al.}(2020)\citenamefont {Komori},
  \citenamefont {Enomoto}, \citenamefont {Ooi}, \citenamefont {Miyazaki},
  \citenamefont {Matsumoto}, \citenamefont {Sudhir}, \citenamefont
  {Michimura},\ and\ \citenamefont {Ando}}]{komori_attonewton_2020}%
  \BibitemOpen
  \bibfield  {author} {\bibinfo {author} {\bibfnamefont {K.}~\bibnamefont
  {Komori}}, \bibinfo {author} {\bibfnamefont {Y.}~\bibnamefont {Enomoto}},
  \bibinfo {author} {\bibfnamefont {C.~P.}\ \bibnamefont {Ooi}}, \bibinfo
  {author} {\bibfnamefont {Y.}~\bibnamefont {Miyazaki}}, \bibinfo {author}
  {\bibfnamefont {N.}~\bibnamefont {Matsumoto}}, \bibinfo {author}
  {\bibfnamefont {V.}~\bibnamefont {Sudhir}}, \bibinfo {author} {\bibfnamefont
  {Y.}~\bibnamefont {Michimura}},\ and\ \bibinfo {author} {\bibfnamefont
  {M.}~\bibnamefont {Ando}},\ }\bibfield  {title} {\bibinfo {title}
  {Attonewton-meter torque sensing with a macroscopic optomechanical torsion
  pendulum},\ }\href {https://doi.org/10.1103/PhysRevA.101.011802} {\bibfield
  {journal} {\bibinfo  {journal} {Phys. Rev. A}\ }\textbf {\bibinfo {volume}
  {101}},\ \bibinfo {pages} {011802} (\bibinfo {year} {2020})}\BibitemShut
  {NoStop}%
\bibitem [{\citenamefont {Braginsky}\ \emph {et~al.}(1992)\citenamefont
  {Braginsky}, \citenamefont {Khalili},\ and\ \citenamefont
  {Thorne}}]{braginsky_khalili_thorne_1992}%
  \BibitemOpen
  \bibfield  {author} {\bibinfo {author} {\bibfnamefont {V.~B.}\ \bibnamefont
  {Braginsky}}, \bibinfo {author} {\bibfnamefont {F.~Y.}\ \bibnamefont
  {Khalili}},\ and\ \bibinfo {author} {\bibfnamefont {K.~S.}\ \bibnamefont
  {Thorne}},\ }\href {https://doi.org/10.1017/CBO9780511622748} {\emph
  {\bibinfo {title} {Quantum Measurement}}}\ (\bibinfo  {publisher} {Cambridge
  University Press},\ \bibinfo {year} {1992})\BibitemShut {NoStop}%
\bibitem [{\citenamefont {González}\ and\ \citenamefont
  {Saulson}(1995)}]{torsion_pendulum}%
  \BibitemOpen
  \bibfield  {author} {\bibinfo {author} {\bibfnamefont {G.~I.}\ \bibnamefont
  {González}}\ and\ \bibinfo {author} {\bibfnamefont {P.~R.}\ \bibnamefont
  {Saulson}},\ }\bibfield  {title} {\bibinfo {title} {Brownian motion of a
  torsion pendulum with internal friction},\ }\href
  {https://doi.org/https://doi.org/10.1016/0375-9601(95)00194-8} {\bibfield
  {journal} {\bibinfo  {journal} {Physics Letters A}\ }\textbf {\bibinfo
  {volume} {201}},\ \bibinfo {pages} {12} (\bibinfo {year} {1995})}\BibitemShut
  {NoStop}%
\bibitem [{Note3()}]{Note3}%
  \BibitemOpen
  \bibinfo {note} {To find spectra of the form of Eqs.~\protect \eqref
  {Eq:general_spectra}, one assumes the free-mass and high-$Q$ limits where the
  mechanical susceptibility behaves as $1/\Omega ^2$; additionally, one must be
  in frequency ranges where $S_{n_F}$ and $S_{n_X}$ are constant polynomial
  roll-off -- typical at frequencies above the resonance of suspended
  oscillators.}\BibitemShut {Stop}%
\bibitem [{\citenamefont {Abbott}\ \emph {et~al.}(2020)\citenamefont {Abbott}
  \emph {et~al.}}]{Guide_to_LIGO_Virgo_detector_noise}%
  \BibitemOpen
  \bibfield  {author} {\bibinfo {author} {\bibfnamefont {B.~P.}\ \bibnamefont
  {Abbott}} \emph {et~al.},\ }\bibfield  {title} {\bibinfo {title} {A guide to
  {LIGO}{\textendash}virgo detector noise and extraction of transient
  gravitational-wave signals},\ }\href
  {https://doi.org/10.1088/1361-6382/ab685e} {\bibfield  {journal} {\bibinfo
  {journal} {Classical and Quantum Gravity}\ }\textbf {\bibinfo {volume}
  {37}},\ \bibinfo {pages} {055002} (\bibinfo {year} {2020})}\BibitemShut
  {NoStop}%
\bibitem [{\citenamefont {Martynov}\ \emph {et~al.}(2016)\citenamefont
  {Martynov}, \citenamefont {Hall} \emph
  {et~al.}}]{Sensitivity_of_Advanced_LIGO}%
  \BibitemOpen
  \bibfield  {author} {\bibinfo {author} {\bibfnamefont {D.~V.}\ \bibnamefont
  {Martynov}}, \bibinfo {author} {\bibnamefont {Hall}}, \emph {et~al.},\
  }\bibfield  {title} {\bibinfo {title} {Sensitivity of the advanced ligo
  detectors at the beginning of gravitational wave astronomy},\ }\href
  {https://doi.org/10.1103/PhysRevD.93.112004} {\bibfield  {journal} {\bibinfo
  {journal} {Phys. Rev. D}\ }\textbf {\bibinfo {volume} {93}},\ \bibinfo
  {pages} {112004} (\bibinfo {year} {2016})}\BibitemShut {NoStop}%
\bibitem [{\citenamefont {Aston}\ \emph {et~al.}(2012)\citenamefont {Aston}
  \emph {et~al.}}]{quadruple_pendulum}%
  \BibitemOpen
  \bibfield  {author} {\bibinfo {author} {\bibfnamefont {S.~M.}\ \bibnamefont
  {Aston}} \emph {et~al.},\ }\bibfield  {title} {\bibinfo {title} {Update on
  quadruple suspension design for advanced ligo},\ }\href
  {https://doi.org/10.1088/0264-9381/29/23/235004} {\bibfield  {journal}
  {\bibinfo  {journal} {Classical and Quantum Gravity}\ }\textbf {\bibinfo
  {volume} {29}},\ \bibinfo {pages} {235004} (\bibinfo {year}
  {2012})}\BibitemShut {NoStop}%
\bibitem [{\citenamefont {Liu}\ and\ \citenamefont
  {Thorne}(2000)}]{coating_thermal}%
  \BibitemOpen
  \bibfield  {author} {\bibinfo {author} {\bibfnamefont {Y.~T.}\ \bibnamefont
  {Liu}}\ and\ \bibinfo {author} {\bibfnamefont {K.~S.}\ \bibnamefont
  {Thorne}},\ }\bibfield  {title} {\bibinfo {title} {Thermoelastic noise and
  homogeneous thermal noise in finite sized gravitational-wave test masses},\
  }\href {https://doi.org/10.1103/PhysRevD.62.122002} {\bibfield  {journal}
  {\bibinfo  {journal} {Phys. Rev. D}\ }\textbf {\bibinfo {volume} {62}},\
  \bibinfo {pages} {122002} (\bibinfo {year} {2000})}\BibitemShut {NoStop}%
\bibitem [{\citenamefont {Müller-Ebhardt}\ \emph {et~al.}(2009)\citenamefont
  {Müller-Ebhardt}, \citenamefont {Rehbein}, \citenamefont {Li}, \citenamefont
  {Mino}, \citenamefont {Somiya}, \citenamefont {Schnabel}, \citenamefont
  {Danzmann},\ and\ \citenamefont
  {Chen}}]{Muller_Ebhardt_quantum_state_preparation}%
  \BibitemOpen
  \bibfield  {author} {\bibinfo {author} {\bibfnamefont {H.}~\bibnamefont
  {Müller-Ebhardt}}, \bibinfo {author} {\bibfnamefont {H.}~\bibnamefont
  {Rehbein}}, \bibinfo {author} {\bibfnamefont {C.}~\bibnamefont {Li}},
  \bibinfo {author} {\bibfnamefont {Y.}~\bibnamefont {Mino}}, \bibinfo {author}
  {\bibfnamefont {K.}~\bibnamefont {Somiya}}, \bibinfo {author} {\bibfnamefont
  {R.}~\bibnamefont {Schnabel}}, \bibinfo {author} {\bibfnamefont
  {K.}~\bibnamefont {Danzmann}},\ and\ \bibinfo {author} {\bibfnamefont
  {Y.}~\bibnamefont {Chen}},\ }\bibfield  {title} {\bibinfo {title}
  {Quantum-state preparation and macroscopic entanglement in gravitational-wave
  detectors},\ }\bibfield  {journal} {\bibinfo  {journal} {Physical Review A}\
  }\textbf {\bibinfo {volume} {80}},\ \href
  {https://doi.org/10.1103/physreva.80.043802} {10.1103/physreva.80.043802}
  (\bibinfo {year} {2009})\BibitemShut {NoStop}%
\bibitem [{\citenamefont {{Rollins}}\ \emph {et~al.}(2020)\citenamefont
  {{Rollins}}, \citenamefont {{Hall}}, \citenamefont {{Wipf}},\ and\
  \citenamefont {{McCuller}}}]{pygwinc}%
  \BibitemOpen
  \bibfield  {author} {\bibinfo {author} {\bibfnamefont {J.~G.}\ \bibnamefont
  {{Rollins}}}, \bibinfo {author} {\bibfnamefont {E.}~\bibnamefont {{Hall}}},
  \bibinfo {author} {\bibfnamefont {C.}~\bibnamefont {{Wipf}}},\ and\ \bibinfo
  {author} {\bibfnamefont {L.}~\bibnamefont {{McCuller}}},\ }\href@noop {}
  {\bibinfo {title} {{pygwinc: Gravitational Wave Interferometer Noise
  Calculator}}},\ \bibinfo {howpublished} {Astrophysics Source Code Library,
  record ascl:2007.020} (\bibinfo {year} {2020}),\ \Eprint
  {https://arxiv.org/abs/2007.020} {ascl:2007.020} \BibitemShut {NoStop}%
\bibitem [{\citenamefont {Wanner}(2019)}]{spacebased}%
  \BibitemOpen
  \bibfield  {author} {\bibinfo {author} {\bibfnamefont {G.}~\bibnamefont
  {Wanner}},\ }\bibfield  {title} {\bibinfo {title} {Space-based gravitational
  wave detection and how lisa pathfinder successfully paved the way},\ }\href
  {https://doi.org/10.1038/s41567-019-0462-3} {\bibfield  {journal} {\bibinfo
  {journal} {Nature Physics}\ }\textbf {\bibinfo {volume} {15}},\ \bibinfo
  {pages} {200} (\bibinfo {year} {2019})}\BibitemShut {NoStop}%
\bibitem [{\citenamefont {Gr{\"o}blacher}\ \emph {et~al.}(2009)\citenamefont
  {Gr{\"o}blacher}, \citenamefont {Hertzberg}, \citenamefont {Vanner},
  \citenamefont {Cole}, \citenamefont {Gigan}, \citenamefont {Schwab},\ and\
  \citenamefont {Aspelmeyer}}]{Gröblacher2009}%
  \BibitemOpen
  \bibfield  {author} {\bibinfo {author} {\bibfnamefont {S.}~\bibnamefont
  {Gr{\"o}blacher}}, \bibinfo {author} {\bibfnamefont {J.~B.}\ \bibnamefont
  {Hertzberg}}, \bibinfo {author} {\bibfnamefont {M.~R.}\ \bibnamefont
  {Vanner}}, \bibinfo {author} {\bibfnamefont {G.~D.}\ \bibnamefont {Cole}},
  \bibinfo {author} {\bibfnamefont {S.}~\bibnamefont {Gigan}}, \bibinfo
  {author} {\bibfnamefont {K.~C.}\ \bibnamefont {Schwab}},\ and\ \bibinfo
  {author} {\bibfnamefont {M.}~\bibnamefont {Aspelmeyer}},\ }\bibfield  {title}
  {\bibinfo {title} {Demonstration of an ultracold micro-optomechanical
  oscillator in a cryogenic cavity},\ }\href
  {https://doi.org/10.1038/nphys1301} {\bibfield  {journal} {\bibinfo
  {journal} {Nature Physics}\ }\textbf {\bibinfo {volume} {5}},\ \bibinfo
  {pages} {485} (\bibinfo {year} {2009})}\BibitemShut {NoStop}%
\bibitem [{\citenamefont {Meng}\ \emph {et~al.}(2022)\citenamefont {Meng},
  \citenamefont {Brawley}, \citenamefont {Khademi}, \citenamefont {Bridge},
  \citenamefont {Bennett},\ and\ \citenamefont
  {Bowen}}]{meng_measurement_2022}%
  \BibitemOpen
  \bibfield  {author} {\bibinfo {author} {\bibfnamefont {C.}~\bibnamefont
  {Meng}}, \bibinfo {author} {\bibfnamefont {G.~A.}\ \bibnamefont {Brawley}},
  \bibinfo {author} {\bibfnamefont {S.}~\bibnamefont {Khademi}}, \bibinfo
  {author} {\bibfnamefont {E.~M.}\ \bibnamefont {Bridge}}, \bibinfo {author}
  {\bibfnamefont {J.~S.}\ \bibnamefont {Bennett}},\ and\ \bibinfo {author}
  {\bibfnamefont {W.~P.}\ \bibnamefont {Bowen}},\ }\bibfield  {title} {\bibinfo
  {title} {Measurement-based preparation of multimode mechanical states},\
  }\href {https://doi.org/10.1126/sciadv.abm7585} {\bibfield  {journal}
  {\bibinfo  {journal} {Science Advances}\ }\textbf {\bibinfo {volume} {8}},\
  \bibinfo {pages} {eabm7585} (\bibinfo {year} {2022})},\ \Eprint
  {https://arxiv.org/abs/https://www.science.org/doi/pdf/10.1126/sciadv.abm7585}
  {https://www.science.org/doi/pdf/10.1126/sciadv.abm7585} \BibitemShut
  {NoStop}%
\end{thebibliography}

%

\end{document}